\documentclass[tighten, twocolumn]{aastex63} 
\usepackage{xcolor, fontawesome, microtype, amsmath}
\definecolor{twitterblue}{RGB}{64,153,255}
\newcommand{\twitter}[1]{\href{https://twitter.com/#1}{\textcolor{twitterblue}{\faTwitter}\,\tt \textcolor{twitterblue}{@#1}}}
\shorttitle{Searching for interstellar quantum communications}
\shortauthors{Michael Hippke}
\begin{document}
\title{Searching for interstellar quantum communications}
\author[0000-0002-0794-6339]{Michael Hippke}
\affiliation{Sonneberg Observatory, Sternwartestr. 32, 96515 Sonneberg, Germany \twitter{hippke}}
\affiliation{Visiting Scholar, Breakthrough Listen Group, Berkeley SETI Research Center, Astronomy Department, UC Berkeley}
\email{michael@hippke.org}

%

\begin{abstract}
The modern search for extraterrestrial intelligence (SETI) began with the seminal publications of \citet{1959Natur.184..844C} and \citet{1961Natur.190..205S}, who proposed to search for narrow-band signals in the radio spectrum, and for optical laser pulses. Over the last six decades, more than one hundred dedicated search programs have targeted these wavelengths; all with null results. All of these campaigns searched for classical communications, that is, for a significant number of photons above a noise threshold; with the assumption of a pattern encoded in time and/or frequency space. I argue that future searches should also target \textit{quantum communications}. They are preferred over classical communications with regards to security and information efficiency, and they would have escaped detection in all previous searches. The measurement of Fock state photons or squeezed light would indicate the artificiality of a signal. I show that quantum coherence is feasible over interstellar distances, and explain for the first time how astronomers can search for quantum transmissions sent by ETI to Earth, using commercially available telescopes and receiver equipment.
\end{abstract}

\keywords{general: extraterrestrial intelligence, instrumentation: detectors, methods: data analysis}

\section{Introduction}
The search for extraterrestrial intelligence (SETI) started in earnest with a discussion between Phil Morrison and his collaborator: ``And Cocconi came to me one day and said, 'You know what, Phil? If there are people out there, won’t they communicate with gamma rays that’ll cross the whole galaxy?' And I said, ’Gee, I know nothing about that (...) Why not use radio? It’s much cheaper. You get many more photons per watt and that must be what counts.’ And we began working on that and pretty soon we knew enough about radio astronomy to publish a paper called ’Searching for Interstellar Communications.’'' \citep{Gingerich2003}. Morrison’s reason for radio, and against optical and $\gamma$-ray communication was that they ``demand very complicated techniques''. However, only months later, the laser was invented \citep{1960Natur.187..493M}. It was clear right away that the tighter beam of optical communication can increase the data rate compared to radio communications \citep{1961Natur.190..205S}, who write that ``No such device was known a year ago''.

Since the 1960s, the vast majority of all SETI experiments have been done in the microwave regime. These search for narrow-band (Hz) microwave (GHz) signals \citep[e.g.,][]{2020AJ....159...86P}, and take into account some amount of the unknown Doppler drift, typically $\pm20$\,Hz\,s$^{-1}$ \citep{2020AJ....160...29S}.

A smaller number of searches has been made in the optical.
In the time domain, the signal assumption is a pulsed laser. Signal photons are expected to arrive in a pulse, e.g., 100 photons per square meter in one 1\,ns bin, whereas starlight noise (with Poisson distribution) would only contribute few photons per nanosecond per square meter \citep[e.g.,][]{ 2004ApJ...613.1270H,2009AsBio...9..345H,2016ApJ...818L..33A}.
In the frequency domain, the signal assumption is a continuous-wave narrow-band laser. Sun-like stars exhibit few natural spectral emission lines only at certain known wavelengths. Additional spectral emission lines, perhaps coinciding with known laser lines, could be artificial \citep[e.g.,][]{2002PASP..114..416R,2017AJ....153..251T,2021arXiv210201910M}.

The general search assumption is that of a beacon, i.e. a lighthouse in the sky: easy to find, but bearing little to no information. Isotropic beacons are most plausible in the microwave regime, because each photon is cheap, and large apertures can be made at low cost. For communications which aim to maximize the amount of data, however, shorter wavelengths are preferred because of their reduced free-space loss \citep{2017arXiv171105761H}. Such collimated beams, however, are rarely intercepted accidentally. Instead, it is much more likely to find signals deliberately directed at the Earth \citep{2014JBIS...67..232F}. They may originate from ETI on distant exoplanets, or from probes or repeater stations in the solar system neighborhood \citep{2020AJ....159...85H}. As humans are already actively communicating into space with radio and laser beams, it is insightful to examine how our own technology tackles interplanetary communications. Consequently, I will discuss examples of and proposals for human communications, under the assumption that some of these methods are replicated by putative ETI.

We are looking (and should keep looking) for narrow-band lighthouse blasts, even though we have found none yet. At the same time, it is possible to expand our search. What is today recorded as just starlight might in fact include a weaker, underlying signal, based on the foundation of quantum communications. 

The idea of quantum communications is common in the science fiction literature \citep[e.g.,][]{cline2015armada}. Only recently, \citet{2020PhRvD.102f3005B} suggested in a paper published in {\it Physical Review D} that ETI signals could ``come from a quantum communication mode rather than the classical communication modes''. As such, it is the first attempt at identifying and addressing one of
the central problems of this subject. Similarly, \citet{2020IJAsB..19..295G} hypothesized in a paper published in the {\it International Journal of Astrobiology} that ``post-singularity civilizations may not be communicating by electromagnetic waves but rather by quantum entanglement'', although without further explanation.

It is sometimes argued in the hallways of astronomy departments that we ``just have to tune into the right band'' and – voil\`{a}  – will be connected to the galactic communication channel. I make this hypothesis testable for starlight and quantum communications.

\section{The case for quantum communications}
\label{sec:method}
I can offer four arguments why ETI will choose to transmit quantum communications instead of simple classical beacons: gate-keeping, supremacy, security, and information efficiency.

\subsection{Gate-keeping}
\label{sec:gatekeeping}
ETI may deliberately choose to make communications invisible for less advanced civilizations. Perhaps most or all advanced civilization feel the need to keep the ``monkeys'' out of the galactic channel, and let members only participate above a certain technological minimum. Mastering quantum communications may reflect this limit. The argument was first made by \citet{1979AcAau...6..213S} when discussing communication by neutrino beams, which he judged as ``so difficult that an advanced civilization may purposely choose such a system in order to find and communicate only with ETCs at their own level of development''. Yet, only 33 years after Subotowicz, Neutrino communications have now been shown at a basic level (0.1\,bits/s) in a practical experiment \citep{2012MPLA...2750077S}. However, communication with Neutrinos has a multitude of practical disadvantages over photons, which makes them unattractive for any conceivable rational being \citep{2018AcAau.151...53H}. 
In practice, quantum Neutrino communication is unrealistically difficult due to their small cross-section, which makes it virtually impossible to detect entangled pairs \citep{2009EL.....8550002B}. Other particles like the neutron are easier to detect, but prohibitively energetically costly to focus into tight beams \citep[Table~1 in][]{2018AcAau.151...53H}. It appears that photons are most suitable for classical {\it and} quantum interstellar communications. The quantum part, however, may serve as ETI's gate-keeping.

\begin{figure*}
\includegraphics[width=.5\linewidth]{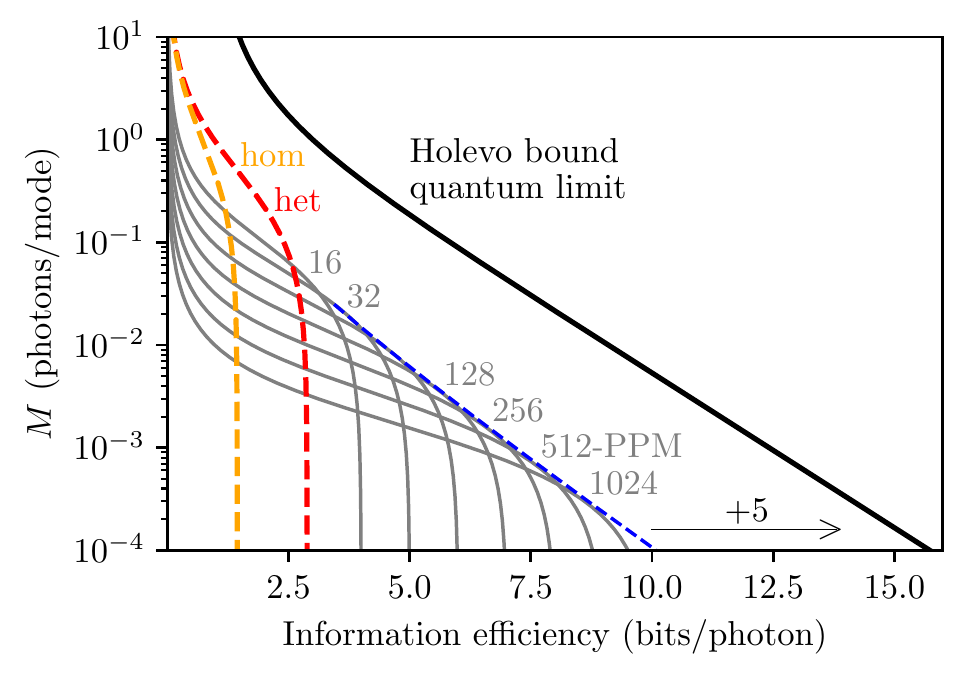}
\includegraphics[width=.5\linewidth]{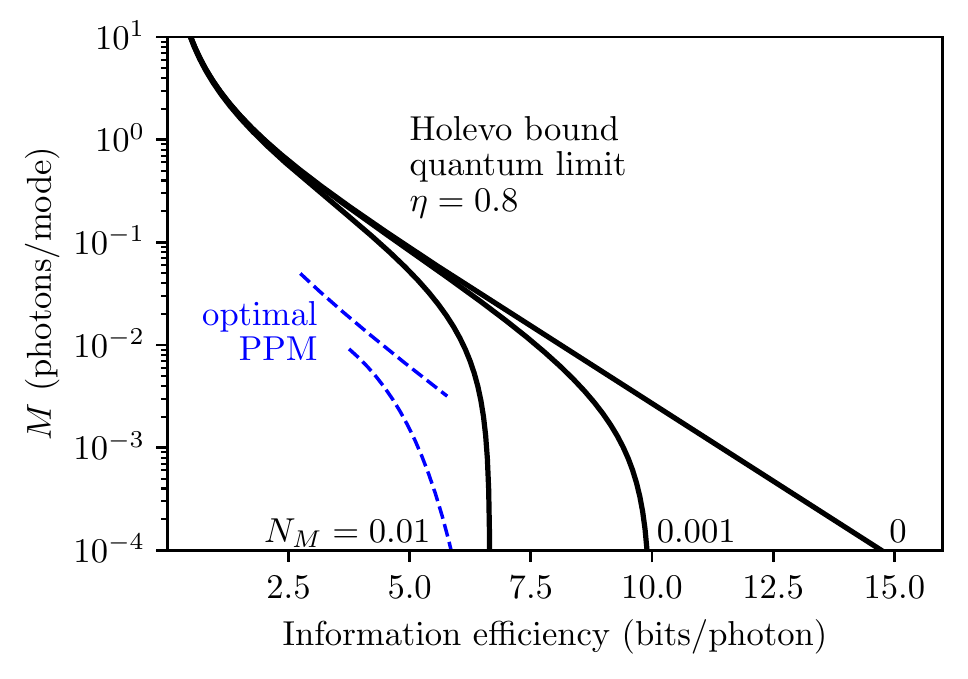}
\caption{\label{fig_h1}Photon information efficiency (PIE, in bits per photon) as a function of dimensional efficiency (DIE, in photons per mode). The area to the right of the Holevo bound, i.e. the quantum limit, is physically forbidden. \textit{Left:} Homodyne (orange dashed) and heterodyne (red dashed) hit brick-wall limits of $\log_2(e) \sim 1.44$ and $2 \log_2(e) \sim 2.89$ bits per photon even at infinite modes (infinite signal-to-noise ratio in classical terms). Pulse-position modulations (PPP) is shown in gray with modulations of different orders. The optimal order is approximated in blue (dashed). The gap in information efficiency between PPM and the quantum limit is about 50\,\%. All curves without noise and losses. \textit{Right:} Holevo limit with receiver efficiency $\eta=0.8$ and different noise levels (in photons per mode). PPM approaches the Holevo limit in the noise-limited case to within $1/\ln(2)\sim1.44$ bits (see text).}
\end{figure*}

\subsection{Quantum supremacy}
Quantum supremacy describes the capability of quantum computers to solve certain problems exponentially faster than classical computers. One example is Shor's algorithm, a method for factoring integers in polynomial time, rather than exponential time. How is this relevant for interstellar communications?

Classical computers were, at their beginning, isolated local objects. Konrad Zuse built the the first Turing-complete computer, the Z3, in his Berlin apartment in 1941. The first computer networks were established about three decades later, beginning with the ARPANET, in 1969. Today's internet connects billions of devices over all of Earth, exchanging large amounts of classical data over classical channels. It is plausible to assume that future space exploration will expand such a network to the planets and the stars \citep{2020AJ....159...85H}.

The rise of quantum computers \citep{2019Natur.574..505A,2020Sci...370.1460Z} paves the way for a similar expansion of the quantum internet \citep{2018arXiv180504360C}. The necessary quantum interconnects can be built with optical interactions of photons and atoms, in order to distribute entanglement across a network and teleport quantum states between nodes \citep{2008Natur.453.1023K}. Connecting quantum computers over quantum networks allows to transmit qubits from one quantum processor to another over long distances. Possible applications include physics and chemistry simulations \citep{1982IJTP...21..467F,1996Sci...273.1073L}, e.g. collider reactions or nitrogen fixation, as well as machine learning, in particular deep learning \citep{2016PhRvA..94b2308B,2020arXiv200300264A}. 

One difficulty for an interstellar quantum network is the delay from light travel time, which is at least several years, compared to a fraction of a second for Earth-based networks. Thus, it is plausible that most quantum computations are performed locally (e.g., the training of a quantum computing deep learning network), and only the condensed qubits are transmitted for remote application (e.g., inference). A second plausible application is optical interferometry for astronomy over arbitrarily long baselines. This is possible by recording the electromagnetic field's amplitude and phase as a function of time. After transmission of a set of qubits to the other's location, both datasets can be combined into one observation \citep{2021arXiv210307590B}.

It may be plausible that future humans, or ETI, build an interstellar quantum network, perhaps for the purpose of distributed quantum computing. This may sound exotic today, but humans have mastered radio telecommunications for only 100 years, and are already at the brink of the transition to quantum communications \citep{2019QS&T....4d5003K,quantum2010014}. After another hundred (or million) years of advancement, it will perhaps appear obvious to any civilization that every reasonable other civilization will have converged on the best solution (which is quantum) in almost no time. After all, it is likely that most civilizations are older and more advanced than ourselves \citep{2020IJAsB..19..430K}.

\subsection{Information security}
A quantum channel can transport {\it classical} information, such as an encyclopedia, as well as {\it quantum} information, i.e., quantum bits (``qubits''). These can be used for secure transmissions with cryptography based on quantum key distribution. We do not know whether ETI values secure interstellar communication, but it is certainly a beneficial tool for expansive civilizations which consist of factions, like humanity today. Therefore, it is plausible that future humans (or ETI) have a desire to implement a secure interstellar network.

For this purpose, quantum key distribution is the method to implements a cryptographic protocol based on quantum mechanics. Its security relies on the foundations of quantum mechanics, which holds as long as the theory is correct. In contrast, traditional public key cryptography relies on the computational difficulty of certain mathematical functions, which may be broken with future technological advances such as quantum computers.

Quantum key distribution enables two distant parties to produce a shared secret key, which can be used to encrypt and decrypt subsequent messages. A unique property of quantum key distribution is the possibility of the two parties to detect a potential third party which intercepts the message. Based on quantum mechanics, the process of measuring a quantum system disturbs it, introducing detectable anomalies. Thus, a communication system can be implemented which detects eavesdropping. If no eavesdropping occurs, the channel is guaranteed to be secure. After key production and distribution, the actual message can be encrypted and transmitted over a classical (or quantum) channel.

The energy cost of secure quantum communications \citep{2017arXiv170508878D}, and the capacity of the quantum-secure channel \citep{2020arXiv200206733G} are active research areas.

\subsection{Information efficiency}
\label{sec:holevo}
A quantum channel can deliver more classical information per unit energy than a classical channel. In other words, to achieve the highest number of bits per photon, i.e. to maximize the data rate, quantum communication is required. Turned the other way around, classical channels are energetically wasteful, because they do not use all information encoding options per photon. 

It must explicitly be stated that there is no exponential gain in photon information efficiency (PIE, in bits per photon) possible through the use of many qubits in superposition. In quantum computing, a memory with $n$ bits of information has $2^n$ possible states. This sometimes leads to the wrong statement that a quantum computer can save $2^n$ bits of information using $n$ quantum logic gates (an exponential gain of quantum over classical). The correct statement is that $n$ qubits can carry up to $2n$ bits of classical information \citep{holevo1973bounds} using super-dense coding \citep{1992PhRvL..69.2881B}.

A detailed summary of classical and quantum capacity, with and without noise, will be published elsewhere (Hippke 2021 in prep.). The interstellar use case is in the photon-starved regime, where the number of modes is much larger than the number of photons, $M \ll 1$. In the low noise case, the capacity of the classical channel is $\log_2 (1/M) - \log_2(\log(1/M))$ bits per photon \citep{Jarzyna2015,2020JLwT...38.2741B}. It can be reached to within a few percent using simple schemes such as high-order pulse-position modulation (PPM). In contrast, the capacity of the quantum channel is $(1+M) \log_2 (1+M)- M \log_2(M) / M + 1$ bits per photon \citet{2004PhRvL..92b7902G}. Thus, the quantum advantage is $ \log_2 \log (1/M) + \log_2(e)$, which is 3--6 bits per photon for $10^{-3}<M<10^{-23}$ (Figure~\ref{fig_h1}). In other words, quantum communications can boost the number of bits per photon by up to $\sim 1/3$. In the noise-limited case, the advantage of quantum communications converges to $1/\ln(2)\sim1.44$ bits per photon \citep{2013PhRvL.110d0501K}.

A quantum advantage of order $1/3$ does not seem like much, but why waste it? It is logical to assume that ETI prefers to transmit more information rather than less, per unit energy. The only quantum ``cost'' is more complicated technology (although this point may be obsolete in a few decades, when quantum technology will have become trivial).

\section{Coherence over interstellar distances}
It was recently shown that quantum coherence of photons over interstellar distances is possible in principle \citep{2020PhRvD.102f3005B}. As a baseline, the largest distance over which successful optical entanglement experiments have been performed on Earth is 144\,km \citep{2009NatPh...5..389F}. The mass density of the Earth's atmosphere at sea level is 1.2\,kg\,m$^{-3}$. Integrated over a column of 144\,km, the column density is $172,800\,$kg\,m$^{-2}$. Satellite-based quantum communications traverse smaller column densities on their combined up and down paths \citep{2019OExpr..2736114Y}. Mass density in space is much lower, but the distances are much greater. Thus, we must examine the column density to the stars in order to determine the feasibility of interstellar quantum communications. This includes interplanetary and interstellar dust (grains) and gas (atoms).

A detailed analysis will be published elsewhere (Hippke 2021 in prep.). In brief, the largest contributors to decoherence are interplanetary dust, mainly produced by mass loss from comets and asteroids, or by fragmentation of meteoroids; it is concentrated mainly in the ecliptic plane. Particles are present in the form of atomic hydrogen, helium, and free electrons; they originate from the solar wind and from interstellar flow. The combined interplanetary and interstellar column density between the Earth and the nearest star, Proxima Centauri, is about $3\times10^{-8}$\,kg\,m$^{-2}$ \citep{2000JGR...10510317M}. This is 8 orders of magnitude less than the 144\,km through Earth's atmosphere. The atmospheric influence on decoherence dominates for all distances $<$kpc and all wavelengths except those in the Lyman continuum ($\approx50-91\,$nm) which is opaque even for the closest stars due to the ionization of neutral hydrogen \citep{2007eua..book.....B}.

\section{Quantum communication implementations}
If quantum is the way to go, what do quantum communications look like? What do we need to search for? There are three cases to consider. Either the transmitter uses quantum building blocks, or the receiver, or both. For quantum cryptography, i.e. secure communications, both ends need to be quantum. To transmit the maximum amount of classical information over a quantum channel, i.e. to achieve the Holevo bound, only one side requires quantum methods \citep{hausladen1996classical,1997PhRvA..56..131S,Holevo1998}. With our current technological understanding, it is unclear which side is easier to implement. If a quantum transmitter is difficult to build, it would be advantageous to use a classical coherent light source (a laser) and let the receiver do the quantum lift. On the other hand, if pair photons can readily be made, e.g., with semiconductor quantum dots \citep{2017PhRvB..95t1410S}, they could be used onboard a spacecraft.

Quantum communications based on photons can be realized with several known physical implementations, or combinations thereof. The first is polarization encoding, where the horizontal and vertical polarization of light is used. The second is through the number of photons (their Fock state), where either an integer number of photons is present, or vacuum. The third option is time-bin encoding, where the time of arrival (early or late) is used. The fourth option is a coherent state of light, known as squeezed light. Here, the information support is quadrature (amplitude-squeezed or phase-squeezed state). Finally, instead of a two-level system, one can use larger dimensional systems such as orbital angular momentum (OAM). At the cost of larger spatial divergence, multiple photons can encode several bits of information using OAM.

In comparison, classical communications in the low $M$ regime most commonly use on-off keying  (or pulse-position modulation in the strong-loss regime), which is the simplest form of amplitude shifts. Simplified, the presence or lack of a carrier (a photon) signifies a binary one or zero. Less commonly, binary digits are encoded on the wavelength of a photon or on its polarization.

\begin{figure*}
\includegraphics[width=\linewidth]{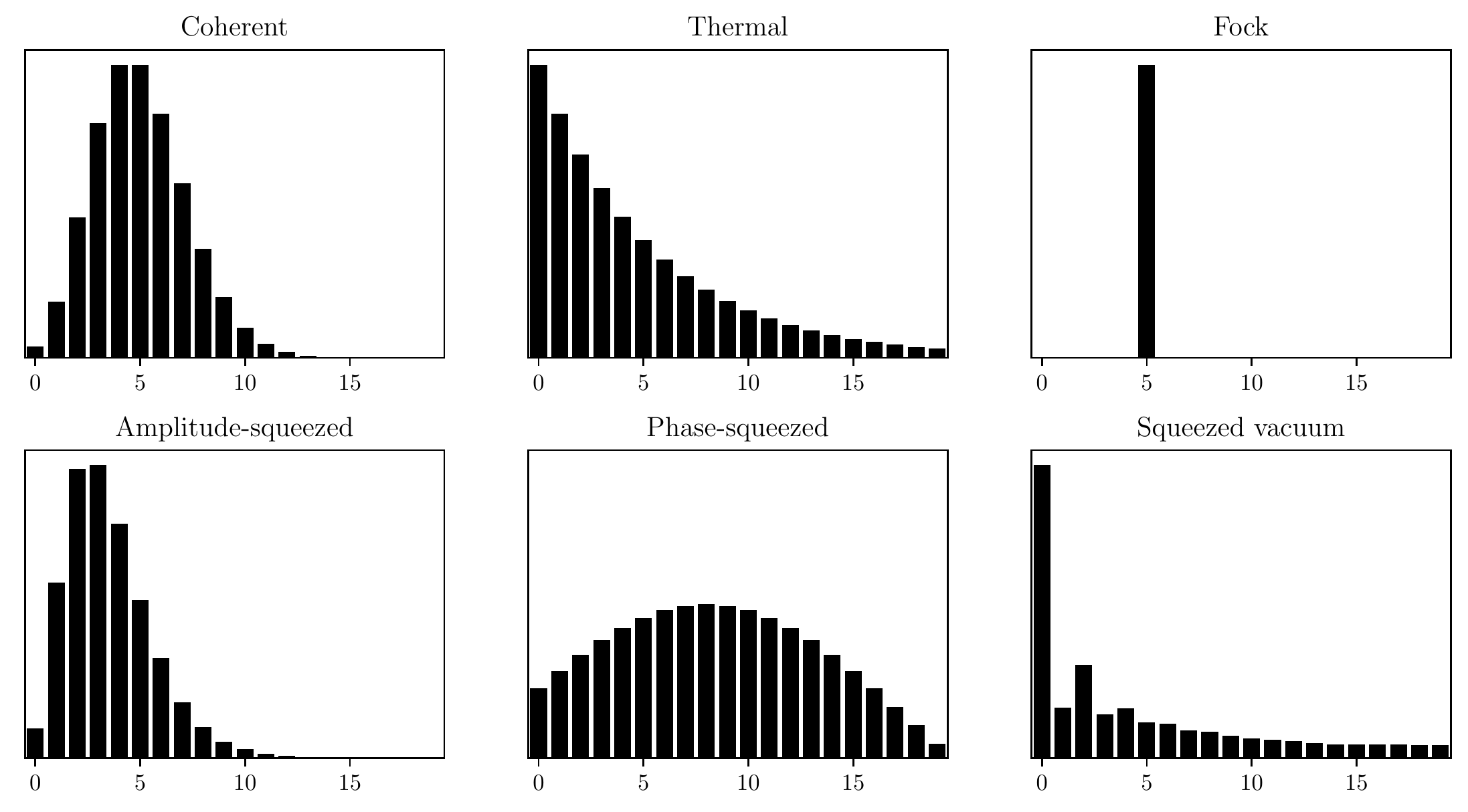}
\caption{\label{fig_squeezing}Photon number distributions for different states of light. The horizontal axis shows the number of photons per unit time, the vertical axis is in arbitrary units. Coherent (laser) light has different temporal bunching compared to thermal light, which is strictly decreasing in the number of photons per unit time. An idealized Fock state with 5 photons is shown in the upper right panel. The amplitude-squeezed state has a denser distribution compared to laser light, while the phase-squeezed state appears wider. The (pure) squeezed-vacuum state is special because is has no contribution from odd-photon-number states.}
\end{figure*}

\subsection{Quantum transmitter}
The first option for a quantum communication scheme is a transmitter using quantum light, paired with a classical receiver. As \citet{2004PhRvL..92b7902G} put it, ``number-state systems for the lossless channel require a non-classical light source, but its receiver can use simple modal photon counting''. This also applies to the lossy bosonic channel; e.g., photons with free-space loss \citep{2012arXiv1202.0518W,2013PhRvL.110u0504O}.

The most common form of nonclassical light is the usage of Fock states: a quantum state with a well-defined number (an exact number) of particles (quanta), here: photons.  A single-mode system in the Fock state $|n\rangle$ contains exactly $n$ excitations of energy, in multiples of Planck's constant $\hbar$. Such a Fock-state photon transmitter could be paired with a classical receiver, counting the specific photon numbers classically, which would achieve the Holevo limit. As the number of photons is affected by quantum uncertainty (following a Poisson distribution), the Fock states improve number state statistics, allowing to surpass the classical Shannon limit. Simplified, classical light comes with a probability in the number of photons per pulse. In a Fock state, this probability is tighter (Figure~\ref{fig_squeezing}), reducing uncertainty for the receiver about what symbol was detected, and thus increasing capacity \citep{Guerrini2019}.

Similar schemes are possible using squeezed light \citep{2018arXiv181109423M}. A useful analogy is Binary Phase Shift Keying (BPSK), where phase-squeezed light provides tighter phase uncertainties \citep{Vourdas1994,2020arXiv200606522F}. From an information perspective, a classical receiver would be sufficient for a perfect measurement, because the heavy lifting is done on the transmitter side. From an OSETI perspective, the measurement of Fock state photons or squeezing would indicate the artificiality of a signal.

\subsection{Quantum receiver}
Alternatively, the transmitter can use classical light, e.g., coherent laser light, while the quantum lifting is done on the receiver end. As \citet{2004PhRvL..92b7902G} put it, a ``coherent-state system uses a classical transmitter, but its detection strategy can be highly non-classical.'' While a classical receiver counts each photons and destroys it during the measurement, a quantum receiver must measure an ensemble of photons jointly before their destruction; the device is thus called a joint-detection receiver. No such device has been built yet, but there exist construction proposals and laboratory experiments.

With a transmitter sending classical laser light, and the receiver performing the quantum uplift, we can already search for such transmissions -- they are ``only'' laser light. When recorded classically (neglecting the quantum receiver), the decoding of the message would remain impossible. The character of the transmission as laser light, however, can prove its artificiality. Still, I will briefly summarize quantum receivers to provide some insights into the future of quantum communication detectors.

At first, it might appear illogical that classical laser light can be paired with an improved (quantum) receiver to generate a higher information efficiency -- after all, where should the additional bits per photon come from? The correct mental image here is that the transmitter indeed sends ``only'' classical light, but with the corresponding classical error probability for each bit. Due to the quantum nature of light, classical communication is not perfect and error-free. Instead, its coding scheme is optimized in such a way that probability of error (for each bit, symbol, message) is below some threshold. Now, a better measurement reduces the measurement errors, allowing for denser coding.

Indeed, some of the first laboratory implementations which surpass the classical Shannon limit are benchmarked with respect to the \citet{Helstrom1976} limit. For example, \citet{2015NaPho...9...48B} describe a receiver which discriminates among four possible  quadrature phase shift keying (QPSK) states, using a 99:1 beamsplitter, local oscillator, and a photon counter. 

Collecting many photons before their destructive measurement is challenging. One method might be a ``slicing receiver'' for Fock-state photons \citep{2013PhRvA..87e2320D} which discriminates between any set of coherent state (laser light) signals using a small all-optical special-purpose quantum computer. The quantum computer must be able to do arbitrary logic on a set of qubits. 

A related scheme uses the concept of Hadamard sequences of input symbols (BPSK at the transmitter side) that are mapped by a linear optical transformation from BPSK onto the PPM format by the quantum receiver \citep{2011PhRvL.106x0502G}. The receiver would need to use quantum memories for collective readout of the classical information \citep{2016JMOp...63.2074K}. Instead of a real quantum memory (which is an area of active research), one could emulate it with known components. To perform collective measurements over multiple time bins requires the synchronization of individual incoming pulses at the receiver, while retaining mutual phase relations. To achieve this with standard tools, one could use fast optical switches and delay lines to equalize pulse arrival times before the Hadamard circuit. However, the whole ensemble grows quickly in size. For a PPM sequence of order 8, the same number of \citet{dolinar1973optimum}-receivers are required, located behind a matrix of $8+7+6..+1=36$ beamsplitters of various transmission coefficients, and a similar number of phase shifters \citep[Figure~2~in][]{2016JMOp...63.2074K}. As explained in section~\ref{sec:holevo}, the interesting regime for interstellar communications lies in the low-$M$ region, requiring large ($>10^3$) PPM orders. Such receivers are impractical to build with physical devices like beamsplitters because of the complexity, size, and transmission losses involved. Instead, quantum computers with quantum memory are preferable.

\subsection{Pre-shared entanglement}
\citet{Bennett2002} proved that pre-shared entanglement increases the capacity of a channel. This scheme is attractive for the use-case of a space probe which launches from the Earth, and takes onboard a box of particles later used to improve data transfer. As explained in the introduction, this is a communication scenario based on human imagination. Communications from ETI may, or may not, leverage similar principles.

First of all, this physically correct and feasible scheme must be distinguished from incorrect descriptions in the literature. Faster-than-light communication using quantum entanglement is a common theme in science fiction \citep[e.g.,][]{cline2015armada}. In some realizations, two distant parties possess a box of entangled particles. They can ``set'' particles in such a way that the other side can then make a measurement, and immediately obtain the information. This is unphysical because faster-than-light signalling results in causality violations which are impossible to accept \citep{1989FoPhL...2..127E,2008IJTP...47.2500B,2012JPhA...45w2001G,2020arXiv200108867F}.

But what about a storage of entangled particles, with light-speed information exchange? Alice sends a box of particles to Bob onboard a spaceship. Bob opens the box and ``modifies'' the quantum states; the entangled partner particle in Alice's box would take the same state, allowing for bits to be communicated -- after some wait time covering for the light travel time. This idea is equally unphysical. It is prohibited by the no-communication theorem \citep{2008PhRvA..77a2113C}: through measurement of an entangled quantum state, it is not possible for one party to communicate information to another party. The fundamental misunderstanding about entanglement is that it only works when measuring the particle's state. If Bob forces his entangled particle into a particular state ($+1$ or $-1$), the entanglement is broken. Alice's subsequent measurement of her previously entangled particle, light years away, will now be $+1$ or $-1$ at random chance.

In contrast, the scheme by \citet{Bennett2002} is correct. Simplified, Alice performs some superoperator on her half, using a pre-shared qubit and a classical bit of data. She then sends the tensor product through the channel, which Bob can use to decode with the help of his half of the pre-shared information. In other words, the pre-shared information can be used to ``compress'' information. The scheme is, in fact, very similar to ``normal'' quantum communication with a combination of entanglement and separately sent classical information \citep[e.g.,][]{2021arXiv210107482H}. The only difference is that the entangled part was shared before the start of the communication.

The gain from pre-sharing is an active research area. In the (theoretical) case of an infinitely large pre-shared cache, and no losses or noise, there is an ``infinite-fold enhancement in communications capacity using pre-shared entanglement'' \citep{2020arXiv200103934G}. Numerically, even an infinite cache only increases the channel capacity (in bits per photon) by $C_E/C = \ln(1/M)$. For $M=10^{-4}$, the increase is at most $\sim 9.2 \times$. In practice, noise and losses eat quickly into this gain and reduce it to a factor of about two or less in all realistic cases.

On a starshot-like probe, the entangled caches would need to withstand decoherence for decades. For this purpose, quantum memory for entangled photons \citep{2011Natur.469..512S} can be used. The current (2021) world-record achieves a six-hour coherence time \citep{2015Natur.517..177Z}, ``made on a ground-state hyperfine transition of europium ion dopants in yttrium orthosilicate using optically detected nuclear magnetic resonance techniques''. Such equipment is not yet suitable to take onboard of a tiny spacecraft, and is in any case insufficient. On the other hand, it is difficult to predict how tomorrow's technology may inhibit decoherence over long times. One advantage for probes is that space is cold and dark, both factors which slow down decoherence.

The concept of \citet{Bennett2002} is a pre-shared entanglement communication protocol applicable to a communication from a spacecraft to its launch point. We may consider the observational signature for the case where Earthlings would intercept such a communication between an alien spacecraft and its home base. This observational signature would be identical to a {\it classical} transmission: The quantum lifting is performed before the two parties separated, and when merging their photons with their pre-shared particles (which we would not possess). Such a communication is not detectable by SETI experiments by its quantum signature. Instead, a SETI experiment would record classical (e.g., laser) communication. Earthlings would lack the required pre-shared entanglement photons to decipher the message. Without them, the classical transmission would appear random in nature. We could only guess that pre-shared entanglement is missing, and that humans are not the intended receiver. That case would imply the existence of at least two communicating parties other than humans.

\section{Signatures and tools}
\label{sec:receiver}
Existing telescope facilities, including those performing optical SETI experiments, are not equipped with quantum communication receivers, because they do not yet exist. However, they often have at their disposal high-quality equipment to detect single photons, take spectra, and perform polarization experiments. I will show that such tools are sufficient to determine the presence of certain quantum communications. Therefore, we can proceed with methods that destroy the actual information content, as long as they provide us with indications of (lost) entanglement. A practical search will be complicated by the fact that we can not know the real characteristics of ETI signals, such as the actual encoding mode and the wavelength. Thus, SETI scientists will have to try many things, just as with today's searches for classical signals. This opens up a whole new field of \textit{Quantum SETI}.

\begin{table*}
\center
\caption{Pulsed OSETI detectors with published sensitivity limits}
\label{tab:previous_obs}
\begin{tabular}{lcccr}
\hline
Obervatory & Cadence (ns) & $\lambda$ (nm) & Sensitivity ($\gamma$\,m$^{-2}\,$ns$^{-1}$) & Reference \\
\hline
Lick 1\,m Nickel               & 5  & $450\dots850$  & 51  & \citet{2001SPIE.4273..173W} \\
Harvard Oak Ridge 1.5\,m       & 5  & $450\dots650$  & 100 & \citet{2004ApJ...613.1270H} \\
Princeton Fitz Randolph 0.9\,m & 5  & $450\dots850$  & 80  & \citet{2004ApJ...613.1270H} \\
STACEE heliostats              & 12 & $300\dots600$  & 10  & \citet{2009AsBio...9..345H} \\
Leuschner 0.8\,m               & 5  & $300\dots700$  & 41  & \citet{2011SPIE.8152E..12K} \\
Harvard 1.8\,m                 & 5  & $300\dots800$  & 60  & \citet{2013PhDT.......161M} \\
Lick 1\,m NIROSETI             & 1  & $950\dots1650$ & 40  & \citet{2014SPIE.9147E..4KM} \\
Veritas 12\,m                  & 50 & $300\dots500$  & 1   & \citet{2016ApJ...818L..33A} \\
Boquete 0.5\,m                 & 5  & $350\dots600$  & 67  & \citet{2016ApJ...825L...5S} \\
\hline
\end{tabular}
\end{table*}

\subsection{Limitations of classical optical SETI}
\label{oseti-limits}
This section proves that current optical SETI experiments can only detect beacon-type signals, but miss communication links weaker in intensity than stars of the first magnitude. I will then describe how to detect such communication links by their quantum effects.

Standard OSETI experiments have a typical minimum sensitivity of order tens of photons per square meter per nanosecond \citep{2019AJ....158..203M} for cadences of order nanosecond (Table~\ref{tab:previous_obs}). Depending on the point of view, this limit can be regarded as astonishingly tight, or terribly loose. It is loose when comparing it to the flux from a Sun-like (G2V) isotropic radiator,

\begin{equation}
F \sim 32 \,{\rm ns}^{-1}\,{\rm m}^{-2} \left(\frac{L}{L_{\odot}}\right) 
\left(\frac{1\,{\rm pc}}{d}\right)^{2}.
\end{equation}

Nominally, when pointing classical OSETI experiments even at bright stars, they will detect nothing, despite many photons hitting the detector (millions per second). On the other hand, when using tight transmitter beams, tight flux limits can be obtained. With a transmitter aperture of $A_t=10\,$m, a receiver $A_r=1\,$m, and $\lambda=1\,\mu$m we get a flux of \citep{2017arXiv171105761H}

\begin{equation}
\label{eq:F}
F \sim 
\left(\frac{1\,{\rm pc}}{d} \right)^{2} 
\left(\frac{A_{\rm t}}{1\,{\rm m}} \right)^{2}
\left(\frac{A_{\rm r}}{1\,{\rm m}} \right)^{2}
\left(\frac{1\,\mu{\rm m}}{\lambda} \right)
\left(\frac{P}{1\,{\rm kW}} \right)
\,\,{\rm (s^{-1})}.
\end{equation}

With a phased (unfilled) array of $A_r=1{,}000\,$m and $P=30\,{\rm MW}$, the received flux is sufficient to outshine Alpha Cen continuously. A km-sized phased aperture was proposed for ``Breakthrough Starshot'' \citep{2016JBIS...69...40L,2016Sci...352.1040M,2017Natur.542...20P} to propel a 1\,g ``space-chip'' with 100\,GW of laser power to $v=0.2\,$c in minutes. Aimed at Alpha Cen, this beamer would appear as a star of magnitude $-4$ on its sky, comparable to Venus and visible in daylight \citep{2016SPIE.9981E..0HL}.
Even a $P=30\,{\rm MW}$ beam, undetected by OSETI experiments, can encode a lot of data. As an example, consider a perfect optical receiver at the time-bandwidth limit, e.g., a photon counter with $3\times10^{14}\,$Hz. At a flux rate of $32\times10^9$\,s$^{-1}$, the number of photons per mode is $M=10^{-4}$. From the Holevo limit, we find ${\rm PIE}=\log_2 (1/M) - \log_2(\log(1/M)) \sim 8$ bits per photon. With a 8192-ary PPM-type modulation, the frame duration would be $3\times10^{-10}\,$s and contain on average one signal photon. Then, the data rate is bound to about 256\,Gbits/s. The second-order correlation function of the PPM modulation would appear with strong anti-bunching (Figure~\ref{fig_bunching}); I will discuss the detection of such signals in section~\ref{sec:bunching}.

An important issue with any interstellar communication (classical or quantum) is the free-space loss. The stated data rate, based on a PIE value and Equation~\ref{eq:F}, is only achievable for parameters exceeding the minimum requirements. A distant ETI signalling towards Earth, however, can not know our receiver's aperture size. It must compensate for this by guessing a minimum size, and choosing its power level sufficiently high. Classical communication would transmit a series of PPM-pulses, each of e.g., kJ power, worth $10^{23}$ photons. On the receiver end, only a few of these photons would be received. For quantum communications, a protocol which requires the reception of a large fraction of photons would thus be inviable. Only quantum communication methods which can handle a large geometric loss are applicable. One of these could employ squeezed light, where each photon has quantum properties, but the information can be spread over many transmitted photons, as with pulses in PPM-like codes. Squeezed light is available today at 20\,W power levels using parametric down-conversion \citep{2013NaPho...7..613A}.

The discovery situation is not much better for spectral OSETI experiments. The usual search paradigm here is that laser light is very narrow-band; broadened to about 0.006\,nm through atmospheric effects \citep{2021arXiv210201910M}. For Holevo-bounded communication, however, narrow-band is not a valid assumption. For time-bandwidth limited pulses with central wavelength $\lambda_0$ and a spectral width (FWHM) $\Delta \lambda$, the pulse duration is approximately 

\begin{equation}
\Delta t_{\rm min} = \frac{\lambda_0^2}{2\,\Delta \lambda c}
\end{equation}

with $c$ as the speed of light. As an example, a near-infrared laser pulse with $\lambda=1\,\mu$m with a $\Delta t_{\rm min} \sim 3 \times 10^{-14}\,$s requires a 5\,\% bandwidth, i.e. $\Delta \lambda =50\,$nm. Such a laser would be undetected by spectral searches.

This completes the point that OSETI experiments can find beacon-type pulses and laser lines, but no continuous communication signals, even when these are as powerful as first magnitude stars. It also opens the question: how do we detect strong laser-like communication signals?

\subsection{Fast photon counters and time-bin encoding}
\label{sec:time-bin}
The best option to detect time-bandwidth limited optical free-space communications would be to use a femto-second photon counter, which would directly show the coherence of the photon arrival time stamps. Unfortunately, these have not yet been invented. The fastest photon detectors today have a timing jitter of $2.6\,$ps (i.e., $2.6\times10^{-12}\,$s) in the optical, based on superconducting nanowires \citep[SNSPDs,][]{2020NaPho..14..250K}. Technology improved by about one order of magnitude per decade \citep{Beskin1997,2009JMOp...56..299T,2013ExMPS..45...69P}. For comparison, the MANIA experiment used 100\,ns counters in the 1980s \citep{1981psec.book..122S}, improving to a few ns in the 2000s \citep{2001SPIE.4273..173W}. If progress continuous in the same pace, we may possess sufficiently fast detectors in about 20 years. 

Until then, we need to use other methods. An experiment with SNSPDs can explore the limits set by atmospheric turbulence already today: optical and NIR laser pulses are smeared to $\Delta t \approx 10^{-12}\,$s \citep{2018JApA...39...74H}. At the time-bandwidth limit, their spectral width would be $\Delta \lambda \approx 1.5\,$nm. These pulses can already be time-resolved. A useful optical SETI experiment would be to time-tag incoming photons with ps accuracy using precise high time-resolution photon counters \citep{2001ApPhL..79..705G,2013RScI...84d6107P}. Time-based correlations can be determined in post with data analysis, such as Fourier transforms and autocorrelation functions, building on the proposal by \citet{2013AsBio..13..521L} and \citet{2019AcAau.156...92S}. These experiments can test incoming sources (e.g., starlight) for the presence of time-bin encoding, which is one way of encoding the quantum information.

\subsection{Laser warners and phase coherence}
\label{sec:laser-warner}
Defense forces try to solve a problem similar to OSETI. While we assume that ETI uses lasers to send signals across space, lasers on Earth are used by the military for targeting, range finding, and missile control. Laser warner receivers have been developed to detect the threat associated with incoming lasers. They use one or several of the following properties of coherent light: high brightness in a narrow spectral band, tight spatial beams, and phase coherence \citep{2019SPIE11161E..0GB}. In an interstellar setting, where the transmitter is far and small, it appears unresolved. Then, only temporal (pulsed) and spectral (narrow-band) properties remain, together with phase coherence.

To detect the presence of coherent light in the presence of noise (chaotic starlight, atmospheric noise etc.), it was already suggested by \citet{1993SPIE.1867...75K} to employ heterodyne and homodyne detection. A recent proposal by \citet{2019PASP..131g4501B} proposes a method for a photon-counting laser coherence detection system. Based on an asymmetric Mach-Zehnder interferometer, it is equally sensitive to pulsed and continuous signals. For a wide spectral coverage, many sensors are required (e.g., 1024 APD cells for the same number of 1\,nm wide frequency bins). Such a detector will be much more sensitive to the information bearing transmissions described in section~\ref{oseti-limits}, because it drops the requirement of a beacon-type threshold per unit time, and instead integrates over arbitrary durations using Fourier transformation during post-processing.

\begin{figure}
\includegraphics[width=\linewidth]{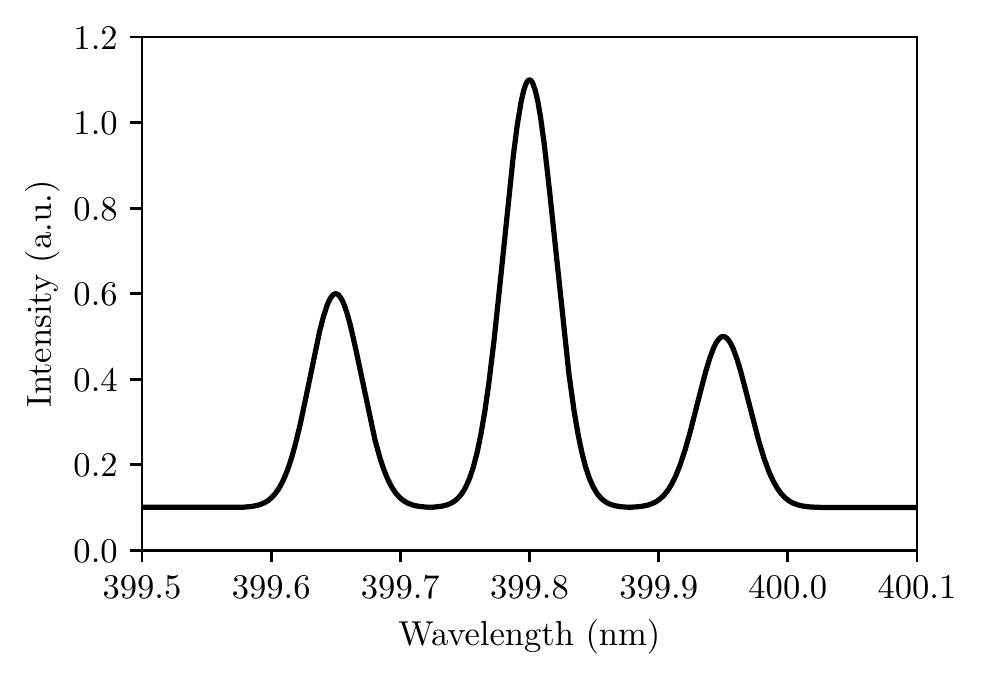}
\caption{\label{fig_triple_pulse}Synthetic spectrum expected from a laser with time-bin encoded qubits. Compared to a classical single frequency signal, three peaks occur, caused by the temporal modulation of the laser. Illustration based on Figure 3 in \citet{2013PhRvL.111o3602D}.}
\end{figure}

\subsection{Spectral encodings}
Classical optical SETI also searches for narrow-band laser lines \citep[e.g.,][]{2017AJ....153..251T}. The assumption is that sun-like stars exhibit few natural spectral emission lines only at certain wavelengths. Thus, after removing known features, any remaining spectral peak could be a laser beam. In these classical searches, the expectation is to find a single peak narrow in frequency, as produced by continuous wave lasers \citep{2002PASP..114..416R}.

Quantum communication signals, however, would be expected to be modulated in order to transport information. Plausible time-bandwidth limits are of order ps and tens of nm (section~\ref{oseti-limits}). Depending on the exact quantum encoding, there could be \textit{three} spectral peaks separated by $\sim1\,$nm and of slightly different amplitude (Figure~\ref{fig_triple_pulse}). Using existing data, one could examine spectra of stars for such multimodal emission features. Instead of a single emission line, quantum optical SETI would search for a frequency comb. Spectra could also be Fourier transformed to search for time-bin encoded signals \citep{2019PASP..131c4502H}.

\begin{figure*}
\includegraphics[width=\linewidth]{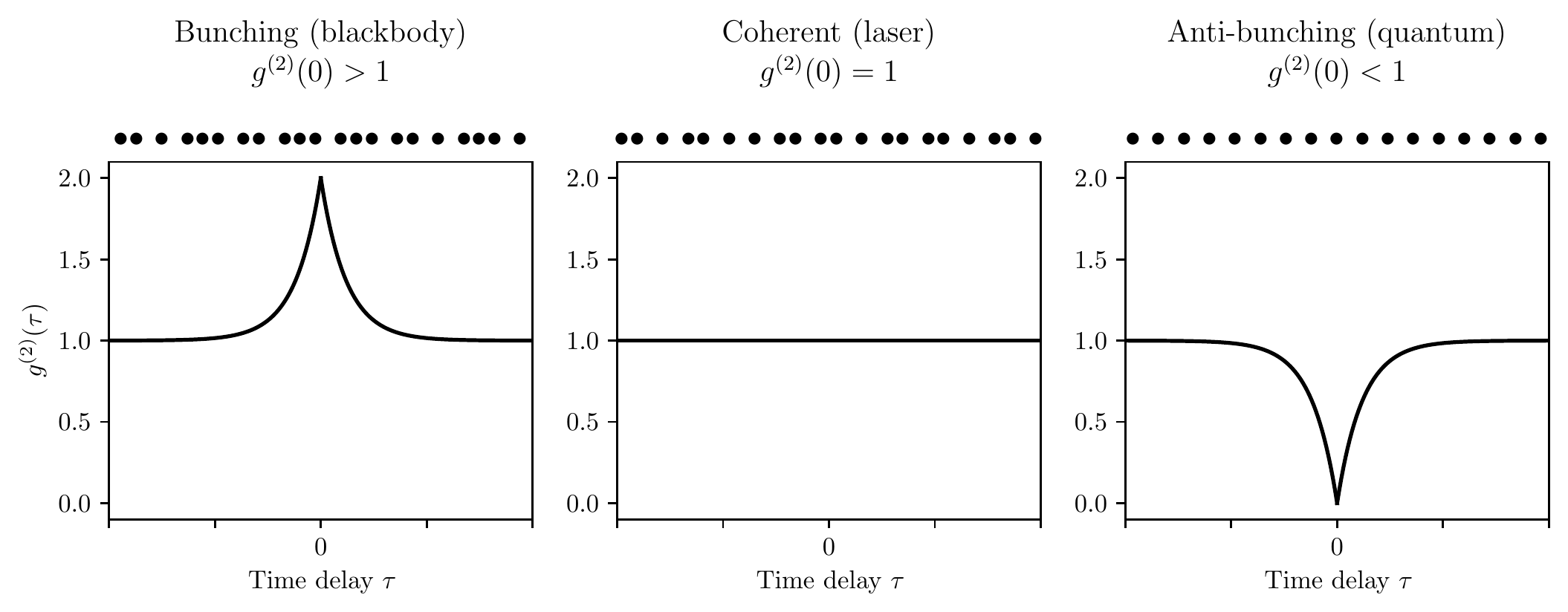}
\caption{\label{fig_bunching}Temporal bunching of photons for super-Poisson distributed chaotic thermal starlight (left), Poisson-distributed coherent laser light (middle) and equally spaced (anti-bunched) quantum light. Cases where $g^{(2)}(0) < 2$ may be  indicative of a mixture of starlight and artificial signals. For $g^{(2)}(0) < 1$ a quantum component is present.}
\end{figure*}

\subsection{Bunching: Hanbury Brown and Twiss effect}
\label{sec:bunching}
Photon (anti-)bunching is an (anti-)temporal correlation in the arrival time of subsequent photons, also known as the Hanbury Brown and Twiss effect \citep{1954PMag...45..663B}. It is leveraged in astronomy for intensity interferometry \citep{1956Natur.178.1046H}. In the original experiment, two telescopes separated by a few meters were aimed at the star Sirius. Each telescope used a photon detector, and a correlation in the arrival time between the two fluctuating intensities was detected. The correlation vanished when the distance between the telescopes was increased. This information can be used to determine the apparent angular size of the star.

There are three regimes of temporal photon correlations. The first is chaotic starlight with a blackbody-like spectrum, which exhibits positive bunching that is super-Poissonian distributed (Figure~\ref{fig_bunching}). In essence, photons arrive bunched (or clustered) in time. With a time difference $\tau$ between subsequent photon detections, one can define the first-order correlation function as $g^{(1)}(\tau)$, and the second order correlation function as

\begin{equation}
g^{(2)}(\tau) = 1 + |g^{(1)}(\tau)|^2.
\end{equation}

Thermal blackbody radiation is the textbook example of $g^{(2)}(\tau)=2$ \citep{1963PhRv..131.2766G,saleh1991fundamentals}. Coherent laser light, i.e. light of a single frequency, shows as $g^{(2)}(\tau)=1$. It follows the well-known Poissonian distribution. Finally, anti-bunching with $g^{(2)}(\tau)<1$ has no classical wave analogue; its distribution is sub-Poissonian. In the limit of perfect anti-bunching, $g^{(2)}(\tau)=0$, the arrival time between photons is constant. As a signal hypothesis, any $g^{(2)}(\tau)<2$ from starlight indicates an artificial component. Anti-bunching ($g^{(2)}(\tau)<1$) would be a strong indication of an information bearing signal based on pulse-position encoding schemes (section~\ref{sec:time-bin} )

In practice, determining $g^{(2)}(\tau)$ for optical light is difficult. The coherence time of optical light near the time-bandwidth limit ($10^{-14}\,$s) is shorter than the best photon counters ($10^{-12}\,$s).
The insufficient timing resolution causes extra noise which requires averaging many signals. Photon bunching was observationally shown for sunlight  with a result of $g^{(2)}(\tau)=1.45 \pm 0.03$ \citep{2014ApJ...789L..10T}, and $g^{(2)}(\tau)=1.7$; lower than the expected value of $2$ due to noise. The same measurement was successfully performed with starlight \citep{2017MNRAS.469.1617T} for stars such as P Cygni with a 1\,m telescope, as well as $\beta$ Canis Majoris and $\epsilon$ Orioni with the Veritas telescope \citep{2020NatAs...4.1164A}. In all cases, a significant value for $g^{(2)}(\tau)>1$ was found. A signal-to-noise ratio discussion for starlight is given in \citet{2019MNRAS.486.5400S}. Similar experiments should be expanded to interesting targets listed in \citet{2020arXiv200611304L}. A measurement of $g^{(2)}(\tau)<1$ would constitute the discovery of quantum communications. The width of the invisible quantum emission line $\Delta \lambda$ (which could shift during the observation, $\lambda_0 \neq \rm{const.}$) can be determined through the relation $\tau=\lambda_0^2 /c\,\Delta \lambda$.

\subsection{Hong-Ou-Mandel interference test}
\label{sec:hong}
Another possibility is the Hong-Ou-Mandel interference test \citep{1987PhRvL..59.2044H}. This experiment is a two-photon interference effect only explainable with quantum optics. It only requires one 50\,\% beamsplitter and two photon counters at its outputs. Chaotic light, e.g., starlight, exhibits distinguishable properties: random polarizations, arrival times, and different wavelengths. These photons will then pass through the beamsplitter randomly and are registered by the detectors. When two distinguishable photons enter the beamsplitter, four possibilities arise with equal probabilities: both photons are transmitted, both photons are reflected, the photon coming in from the first end is reflected and the other transmitted, and the opposite of that. 

In the case a pair of \textit{indistinguishable} photons enters the beamsplitter, a quantum phenomenon occurs. Then, both photons will pass through to one of the sides: one photon counter will count two photons, the other photon counter will count zero photons. While this can also happen by 50\,\% chance for chaotic light, it will \textit{always} happen for indistinguishable photons.

This experiment has been performed with sunlight and light created on the Earth \citep{ 2019PhRvL.123h0401D}. Despite the large physical separation of both sources, when filtered so that the photons become indistinguishable, they exhibit the quantum interference and allow for entanglement tests.

The signal hypothesis for a quantum SETI experiment, based on the Hong-Ou-Mandel interference test, would be as follows. A transmission would consist of a large number of indistinguishable photon pairs. Each pair might be distinguishable from other pairs in frequency, polarization, and arrival time; so that their integrated sum may look natural (e.g., replicate a blackbody spectrum). The signal, however, might partially or totally consist of indistinguishable photon pairs. One method to produce such pair photons are semiconductor quantum dots \citep{2017PhRvB..95t1410S}. An OSETI experiment could use two optical telescopes, focusing their light into each side of a beamsplitter. Through data analysis, one would find a pair ratio exceeding 50\,\% (noise and signal photons combined) in case an artificial signal fraction is the experiment. In practice, the experiment requires very low-noise photon detectors, but is realistically possible at least for the brightest stars, as shown by  \citet{2019PhRvL.123h0401D}.

\begin{figure}
\includegraphics[width=\linewidth]{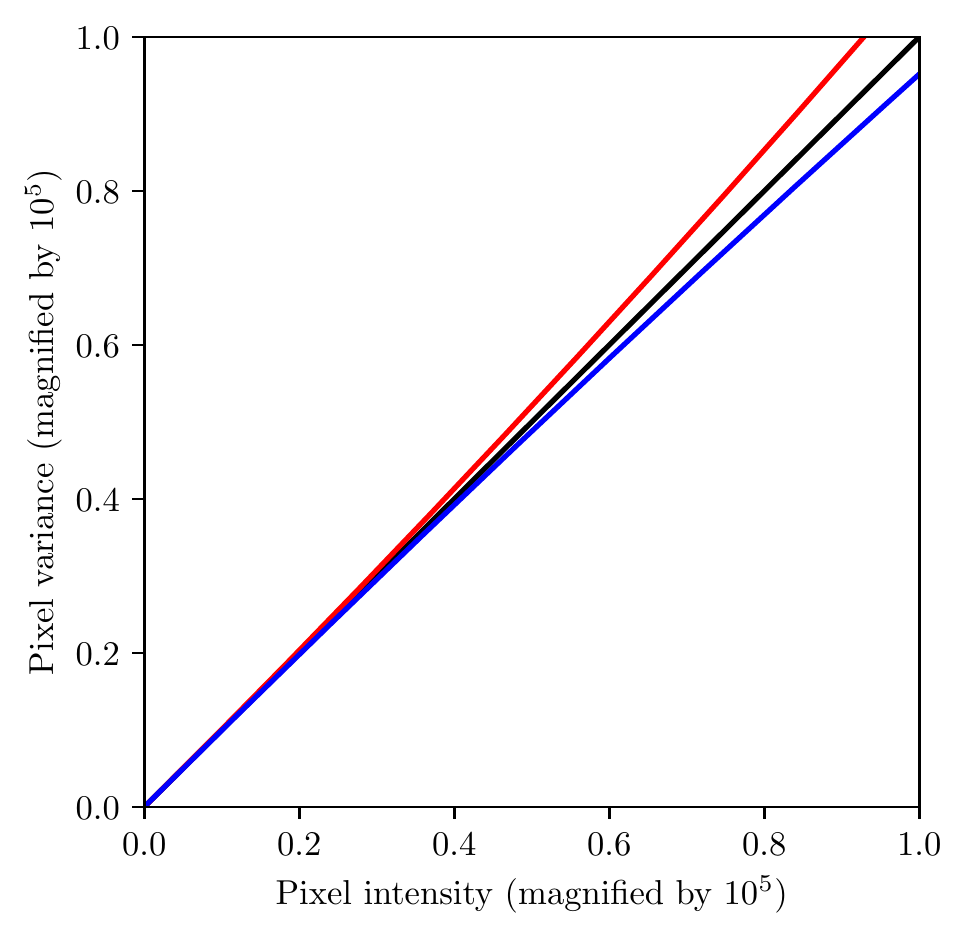}
\caption{\label{fig_squeezing_with_ccd}Signature of squeezed light on a CCD. Pixel variance as a function of pixel intensity for coherent light (laser, black, straight line), squeezed light (red), and anti-squeezed light (blue). Both quadratic functions are shown with coefficients enlarged by $10^5$ for clarity.}
\end{figure}

\subsection{Squeezed light}
\label{sec:squeezed}
Due to the uncertainty principle, a measurement of the amplitude of the light field delivers values within an uncertainty region. Coherent states have circularly symmetric uncertainty regions. The uncertainty relation requires some minimum noise amplitudes for the amplitude and phase. Reducing either amplitude or phase uncertainty, at the expense of the other, constitutes ``squeezed light'' \citep{2017PhR...684....1S}.

First produced by degenerate parametric down conversion in an optical cavity \citep{1986PhRvL..57.2520W}, squeezed light can be made by using optical nonlinear interactions, such as frequency doubling, with the Kerr nonlinearity in optical fibers, or by atom-light interactions. Semiconductor lasers can produce amplitude-squeezed light when operated with a stabilized pump current. A famous application is in quantum metrology \citep{2006PhRvL..96a0401G} by the Laser Interferometer Gravitational-wave Observatory (LIGO), where a sub-threshold optical parametric oscillator produces a squeezed vacuum state \citep{2013NaPho...7..613A}. It allows to improve detector noise performance at certain frequencies by a factor of a few, and makes the detection of gravitational waves from merging black holes, and other sources, possible \citep{2016PhRvL.116f1102A}.

Squeezed light is not expected to occur from natural astrophysical processes. We may assume, however, that ETI signals leverage squeezing to maximize data rate of transmissions. As outlined in section~\ref{sec:holevo}, squeezed light can be used to achieve the Holevo bound of maximum photon information efficiency.

To test the presence of squeezed light, various methods are available. Most commonly, homodyne or heterodyne detection by beating the squeezed light field with a classical local oscillator is used. Other options are direct intensity detection or photon counting (for intensity-squeezed light only), or with a phase-shifting cavity \citep{vogel2006quantum}. Recently, a simpler method using a standard CCD cameras was described, based on the first two moments of the recorded pixel-to-pixel photon fluctuation statistics \citep{2020PhRvL.125k3602M}. Simplified, starlight is expected to produce a linear intensity-variance correlation, assuming that the (CCD) detector is linear in its response. Phase-squeezed (and anti-squeezed light) will however show a quadratic relation between intensity and variance (Figure~\ref{fig_squeezing_with_ccd}). The effect is quite small with a quadratic coefficient of order $10^{-6}$ and thus requires the averaging of many photons. Still, an experiment using the existing CCD detectors on telescopes should be straightforward, because the squeezed-light test is performed using data analysis on the raw sensor data, without any need for additional equipment.

Squeezed light also exhibits second-order correlation effects as for bunched light. \citep[For example, compare Figure~\ref{fig_bunching} with Figure 2 in][]{2012JOSAB..29...15W}. However, I could not find any relevant practical work in this respect. Instead, it is more common to integrate the incoming flux over a certain time period and plot the photon number statistic, i.e., create a histogram of counted photons within this time \citep{1997Natur.387..471B}. Such a histogram can then be fit to a set of distributions e.g., using QuTiP \citep{2013CoPhC.184.1234J}
for thermal and laser light, as well as various quantum states as shown in Figure~\ref{fig_squeezing}.

\subsection{Orbital angular momentum}
Photons can carry a field spatial distribution in the form of orbital angular momentum (OAM). This degree of freedom can be used for quantum communications, such as secure direct communication \citep{2020MPLB...3450017L}. For interstellar communication purposeses, however, POAM is unattractive for two reasons.

First, it also occurs in many natural sources. The presence of astrophysical OAM has been shown using classical receivers on radio telescopes \citep{2005OExpr..13..873T}, detecting twisted light caused by the spin of the M87 black hole \citep{2020MNRAS.492L..22T}. Here, electromagnetic radiation is decomposed from the recorded voltage data into a set of discrete orthogonal eigenmodes. OAM has also been detected as a natural feature of optical starlight \citep{2014A&A...567A.114O} and sunlight \citep{2011A&A...526A..56U}. An optical telescope is equipped with a vortex diffraction grating in front of the CCD, which produces three main OAM diffraction orders.

Second, interstellar distances are very large ($>10^{16}\,$m), so that the free space loss from diffraction, following the inverse square law, is very relevant. With OAM, the beam divergence increases even more, at least with the square root of the number of modes \citep{2015NJPh...17b3011P,2017OExpr..2511265P}. Figure 1 in \citet{2009JOptA..11i4021B} shows the difficulty in measuring spatially unresolved OAM: in the beam centre there is hardly any intensity, and off-axis there is hardly any phase change. Large apertures (or interferometers) are required to make measurements plausible in practice. In any case, using OAM reduces data rates over interstellar distances: beam divergence increases (flux decreases) with $\sqrt{1/M}$, while PIE increases only as $\log_2(1/M)$ bits per photon.

\section{Conclusion and next steps}
Some light from the heavens may be artificial in origin. It is worth exploring all signs of artificiality, including quantum properties. I propose to perform the following tests using available telescopes and equipment.

Squeezed light can be found with a CCD by measuring the first two moments of the recorded pixel-to-pixel photon fluctuation statistics \citep{2020PhRvL.125k3602M}. A test for Fock states can be made by creating photon number statistic, i.e., creating a histogram of counted photons within this time \citep{1997Natur.387..471B}. Time-bin encoded qubits in the frequency domain can be seen as e.g., triplet spectral emission lines caused by the temporal modulation of the laser \citep{2013PhRvL.111o3602D}. Alternatively, the events of high time-resolution photon counters (SNSPDs with ps timing) can be recorded \citep{2001ApPhL..79..705G,2013RScI...84d6107P}, and searched for temporal correlations via data analysis such as Fourier transforms and autocorrelation tests.

Photon bunching can be found through the Hanbury Brown and Twiss effect (section~\ref{sec:bunching}) following the methods from \citet{2014ApJ...789L..10T,2017MNRAS.469.1617T,2020NatAs...4.1164A}. Experiments can use optical telescopes and VERITAS for stars like P Cygni, $\beta$ Canis Majoris and $\epsilon$ Orioni. SNR calculations are provided by \citet{2019MNRAS.486.5400S}. A target list should include Helium white dwarfs \citep{2020IAUGA..30..480F} and targets listed in \citet{2020arXiv200611304L}.

The Hong-Ou-Mandel interference test (section~\ref{sec:hong}) can be performed to search for indistinguishable photons. I propose to check the 100 brightest stars for artificial quantum signals, using twin telescopes with beamsplitters and photon counters. 

Finally, coherence detecting systems of various types, like laser warners, should be pointed towards the stars. One example is the asymmetric Mach-Zehnder interferometer described by \citet{2019PASP..131g4501B}.

It may well be that a search dedicated for quantum signals is also sensitive to classical signals, e.g. in the form of a spread spectrum. The possibility of such a by-catch is not a valid argument against quantum SETI. 

As quoted in the introduction, Giuseppe Cocconi proposed to search for radio signals because ``photons per watt must be what counts''. This is correct for isotropic lighthouse-style beacons. For point-to-point communications in an interstellar network, however, tight beams at e.g., optical frequencies are preferred \citep{2017arXiv171105761H,2020AJ....159...85H}. Then, the {\it number of bits per watt} must be what counts. These are maximized with {\it quantum communications}. 

Following the SETI classification of \citet{2020IJAsB..19..237S}, one important aspect of a potential ETI signal is its level of {\it ambiguity}, i.e. its potential to be a natural phenomenon unrelated to intelligent life. With quantum communications, a detection of Fock state photons or squeezed light would prove the artificiality of a signal, because there is no natural process which could create them. It is also straightforward to search for such signals, which is why we should start looking for them. They may offer the missing access link to an interstellar communication network.

{\it Acknowledgements} I thank Arjun Berera for helpful discussions.

\bibliography{references}

\begin{thebibliography}{}
\expandafter\ifx\csname natexlab\endcsname\relax\def\natexlab#1{#1}\fi
\providecommand{\url}[1]{\href{#1}{#1}}
\providecommand{\dodoi}[1]{}
\providecommand{\doeprint}[1]{}
\providecommand{\doarXiv}[1]{\href{https://arxiv.org/abs/#1}{\nolinkurl{https://arxiv.org/abs/#1}}}

\bibitem[{{Aasi} {et~al.}(2013){Aasi}, {Abadie}, {Abbott}, {Abbott}, {Abbott},
  {Abernathy}, {Adams}, {Adams}, {Addesso}, {Adhikari}, {Affeldt}, {Aguiar},
  {Ajith}, {Allen}, {Amador Ceron}, {Amariutei}, {Anderson}, {Anderson},
  {Arai}, {Araya}, {Arceneaux}, {Ast}, {Aston}, {Atkinson}, {Aufmuth},
  {Aulbert}, {Austin}, {Aylott}, {Babak}, {Baker}, {Ballmer}, {Bao},
  {Barayoga}, {Barker}, {Barr}, {Barsotti}, {Barton}, {Bartos}, {Bassiri},
  {Batch}, {Bauchrowitz}, {Behnke}, {Bell}, {Bell}, {Bergmann}, {Berliner},
  {Bertolini}, {Betzwieser}, {Beveridge}, {Beyersdorf}, {Bhadbhade}, {Bilenko},
  {Billingsley}, {Birch}, {Biscans}, {Black}, {Blackburn}, {Blackburn},
  {Blair}, {Bland}, {Bock}, {Bodiya}, {Bogan}, {Bond}, {Bork}, {Born}, {Bose},
  {Bowers}, {Brady}, {Braginsky}, {Brau}, {Breyer}, {Bridges}, {Brinkmann},
  {Britzger}, {Brooks}, {Brown}, {Brown}, {Buckland}, {Br{\"u}ckner},
  {Buchler}, {Buonanno}, {Burguet-Castell}, {Byer}, {Cadonati}, {Camp},
  {Campsie}, {Cannon}, {Cao}, {Capano}, {Carbone}, {Caride}, {Castiglia},
  {Caudill}, {Cavagli{\`a}}, {Cepeda}, {Chalermsongsak}, {Chao}, {Charlton},
  {Chen}, {Chen}, {Cho}, {Chow}, {Christensen}, {Chu}, {Chua}, {Chung},
  {Ciani}, {Clara}, {Clark}, {Clark}, {Constancio Junior}, {Cook}, {Corbitt},
  {Cordier}, {Cornish}, {Corsi}, {Costa}, {Coughlin}, {Countryman}, {Couvares},
  {Coward}, {Cowart}, {Coyne}, {Craig}, {Creighton}, {Creighton}, {Cumming},
  {Cunningham}, {Dahl}, {Damjanic}, {Danilishin}, {Danzmann}, {Daudert},
  {Daveloza}, {Davies}, {Daw}, {Dayanga}, {Deleeuw}, {Denker}, {Dent},
  {Dergachev}, {Derosa}, {Desalvo}, {Dhurandhar}, {di Palma}, {D{\'\i}az},
  {Dietz}, {Donovan}, {Dooley}, {Doravari}, {Drasco}, {Drever}, {Driggers},
  {Du}, {Dumas}, {Dwyer}, {Eberle}, {Edwards}, {Effler}, {Ehrens},
  {Eikenberry}, {Engel}, {Essick}, {Etzel}, {Evans}, {Evans}, {Evans},
  {Factourovich}, {Fairhurst}, {Fang}, {Farr}, {Farr}, {Favata}, {Fazi},
  {Fehrmann}, {Feldbaum}, {Finn}, {Fisher}, {Foley}, {Forsi}, {Fotopoulos},
  {Frede}, {Frei}, {Frei}, {Freise}, {Frey}, {Fricke}, {Friedrich},
  {Fritschel}, {Frolov}, {Fujimoto}, {Fulda}, {Fyffe}, {Gair}, {Garcia},
  {Gehrels}, {Gelencser}, {Gergely}, {Ghosh}, {Giaime}, {Giampanis},
  {Giardina}, {Gil-Casanova}, {Gill}, {Gleason}, {Goetz}, {Gonz{\'a}lez},
  {Gordon}, {Gorodetsky}, {Gossan}, {Go{\ss}ler}, {Graef}, {Graff}, {Grant},
  {Gras}, {Gray}, {Greenhalgh}, {Gretarsson}, {Griffo}, {Grote}, {Grover},
  {Grunewald}, {Guido}, {Gustafson}, {Gustafson}, {Hammer}, {Hammond}, {Hanks},
  {Hanna}, {Hanson}, {Haris}, {Harms}, {Harry}, {Harry}, {Harstad}, {Hartman},
  {Haughian}, {Hayama}, {Heefner}, {Heintze}, {Hendry}, {Heng}, {Heptonstall},
  {Heurs}, {Hewitson}, {Hild}, {Hoak}, {Hodge}, {Holt}, {Holtrop}, {Hong},
  {Hooper}, {Hough}, {Howell}, {Huang}, {Huerta}, {Hughey}, {Huttner}, {Huynh},
  {Huynh-Dinh}, {Ingram}, {Inta}, {Isogai}, {Ivanov}, {Iyer}, {Izumi},
  {Jacobson}, {James}, {Jang}, {Jang}, {Jesse}, {Johnson}, {Jones}, {Jones},
  {Jones}, {Ju}, {Kalmus}, {Kalogera}, {Kandhasamy}, {Kang}, {Kanner},
  {Kasturi}, {Katsavounidis}, {Katzman}, {Kaufer}, {Kawabe}, {Kawamura},
  {Kawazoe}, {Keitel}, {Kelley}, {Kells}, {Keppel}, {Khalaidovski}, {Khalili},
  {Khazanov}, {Kim}, {Kim}, {Kim}, {Kim}, {Kim}, {King}, {Kinzel}, {Kissel},
  {Klimenko}, {Kline}, {Kokeyama}, {Kondrashov}, {Koranda}, {Korth}, {Kozak},
  {Kozameh}, {Kremin}, {Kringel}, {Krishnan}, {Kucharczyk}, {Kuehn}, {Kumar},
  {Kumar}, {Kuper}, {Kurdyumov}, {Kwee}, {Lam}, {Landry}, {Lantz}, {Lasky},
  {Lawrie}, {Lazzarini}, {Le Roux}, {Leaci}, {Lee}, {Lee}, {Lee}, {Lee},
  {Leong}, {Levine}, {Lhuillier}, {Lin}, {Litvine}, {Liu}, {Liu}, {Lockerbie},
  {Lodhia}, {Loew}, {Logue}, {Lombardi}, {Lormand}, {Lough}, {Lubinski},
  {L{\"u}ck}, {Lundgren}, {MacArthur}, {MacDonald}, {Machenschalk}, {Macinnis},
  {MacLeod}, {Maga{\~n}a-Sandoval}, {Mageswaran}, {Mailand}, {Manca}, {Mandel},
  {Mandic}, {M{\'a}rka}, {M{\'a}rka}, {Markosyan}, {Maros}, {Martin}, {Martin},
  {Martinov}, {Marx}, {Mason}, {Matichard}, {Matone}, {Matzner}, {Mavalvala},
  {May}, {Mazzolo}, {McAuley}, {McCarthy}, {McClelland}, {McGuire}, {McIntyre},
  {McIver}, {Meadors}, {Mehmet}, {Meier}, {Melatos}, {Mendell}, {Mercer},
  {Meshkov}, {Messenger}, {Meyer}, {Miao}, {Miller}, {Mingarelli}, {Mitra},
  {Mitrofanov}, {Mitselmakher}, {Mittleman}, {Moe}, {Mokler}, {Mohapatra},
  {Moraru}, {Moreno}, {Mori}, {Morriss}, {Mossavi}, {Mow-Lowry}, {Mueller},
  {Mueller}, {Mukherjee}, {Mullavey}, {Munch}, {Murphy}, {Murray}, {Mytidis},
  {Nanda Kumar}, {Nash}, {Nayak}, {Necula}, {Newton}, {Nguyen}, {Nishida},
  {Nishizawa}, {Nitz}, {Nolting}, {Normandin}, {Nuttall}, {O'Dell}, {O'Reilly},
  {O'Shaughnessy}, {Ochsner}, {Oelker}, {Ogin}, {Oh}, {Oh}, {Ohme},
  {Oppermann}, {Osthelder}, {Ott}, {Ottaway}, {Ottens}, {Ou}, {Overmier},
  {Owen}, {Padilla}, {Pai}, {Pan}, {Pankow}, {Papa}, {Paris}, {Parkinson},
  {Pedraza}, {Penn}, {Peralta}, {Perreca}, {Phelps}, {Pickenpack}, {Pierro},
  {Pinto}, {Pitkin}, {Pletsch}, {P{\"o}ld}, {Postiglione}, {Poux}, {Predoi},
  {Prestegard}, {Price}, {Prijatelj}, {Privitera}, {Prokhorov}, {Puncken},
  {Quetschke}, {Quintero}, {Quitzow-James}, {Raab}, {Radkins}, {Raffai},
  {Raja}, {Rakhmanov}, {Ramet}, {Raymond}, {Reed}, {Reed}, {Reid}, {Reitze},
  {Riesen}, {Riles}, {Roberts}, {Robertson}, {Robinson}, {Roddy}, {Rodriguez},
  {Rodriguez}, {Rodruck}, {Rollins}, {Romie}, {R{\"o}ver}, {Rowan},
  {R{\"u}diger}, {Ryan}, {Salemi}, {Sammut}, {Sandberg}, {Sanders}, {Sankar},
  {Sannibale}, {Santamar{\'\i}a}, {Santiago-Prieto}, {Santostasi},
  {Sathyaprakash}, {Saulson}, {Savage}, {Schilling}, {Schnabel}, {Schofield},
  {Schuette}, {Schulz}, {Schutz}, {Schwinberg}, {Scott}, {Scott}, {Seifert},
  {Sellers}, {Sengupta}, {Sergeev}, {Shaddock}, {Shahriar}, {Shaltev}, {Shao},
  {Shapiro}, {Shawhan}, {Shoemaker}, {Sidery}, {Siemens}, {Sigg}, {Simakov},
  {Singer}, {Singer}, {Sintes}, {Skelton}, {Slagmolen}, {Slutsky}, {Smith},
  {Smith}, {Smith}, {Smith-Lefebvre}, {Son}, {Sorazu}, {Souradeep}, {Stefszky},
  {Steinert}, {Steinlechner}, {Steinlechner}, {Steplewski}, {Stevens},
  {Stochino}, {Stone}, {Strain}, {Strigin}, {Stroeer}, {Stuver},
  {Summerscales}, {Susmithan}, {Sutton}, {Szeifert}, {Talukder}, {Tanner},
  {Tarabrin}, {Taylor}, {Thomas}, {Thomas}, {Thorne}, {Thorne}, {Thrane},
  {Tiwari}, {Tokmakov}, {Tomlinson}, {Torres}, {Torrie}, {Traylor}, {Tse},
  {Ugolini}, {Unnikrishnan}, {Vahlbruch}, {Vallisneri}, {van der Sluys}, {van
  Veggel}, {Vass}, {Vaulin}, {Vecchio}, {Veitch}, {Veitch}, {Venkateswara},
  {Verma}, {Vincent-Finley}, {Vitale}, {Vo}, {Vorvick}, {Vousden},
  {Vyatchanin}, {Wade}, {Wade}, {Wade}, {Waldman}, {Wallace}, {Wan}, {Wang},
  {Wang}, {Wang}, {Wanner}, {Ward}, {Was}, {Weinert}, {Weinstein}, {Weiss},
  {Welborn}, {Wen}, {Wessels}, {West}, {Westphal}, {Wette}, {Whelan},
  {Whitcomb}, {Wiseman}, {White}, {Whiting}, {Wiesner}, {Wilkinson}, {Willems},
  {Williams}, {Williams}, {Williams}, {Willis}, {Willke}, {Wimmer},
  {Winkelmann}, {Winkler}, {Wipf}, {Wittel}, {Woan}, {Wooley}, {Worden},
  {Yablon}, {Yakushin}, {Yamamoto}, {Yancey}, {Yang}, {Yeaton-Massey},
  {Yoshida}, {Yum}, {Zanolin}, {Zhang}, {Zhang}, {Zhao}, {Zhu}, {Zhu}, {Zotov},
  {Zucker}, \& {Zweizig}}]{2013NaPho...7..613A}
{Aasi}, J., {Abadie}, J., {Abbott}, B.~P., {et~al.} 2013,
  \href{http://dx.doi.org/10.1038/nphoton.2013.177}{\color{magenta}Nature
  Photonics},
  \href{https://ui.adsabs.harvard.edu/abs/2013NaPho...7..613A}{\color{blue}7},
  \href{https://ui.adsabs.harvard.edu/abs/2013NaPho...7..613A}{\color{blue}613}

\bibitem[{{Abbott} {et~al.}(2016){Abbott}, {Abbott}, {Abbott}, {Abernathy},
  {Acernese}, {Ackley}, {Adams}, {Adams}, {Addesso}, {Adhikari}, {Adya},
  {Affeldt}, {Agathos}, {Agatsuma}, {Aggarwal}, {Aguiar}, {Aiello}, {Ain},
  {Ajith}, {Allen}, {Allocca}, {Altin}, {Anderson}, {Anderson}, {Arai},
  {Arain}, {Araya}, {Arceneaux}, {Areeda}, {Arnaud}, {Arun}, {Ascenzi},
  {Ashton}, {Ast}, {Aston}, {Astone}, {Aufmuth}, {Aulbert}, {Babak}, {Bacon},
  {Bader}, {Baker}, {Baldaccini}, {Ballardin}, {Ballmer}, {Barayoga},
  {Barclay}, {Barish}, {Barker}, {Barone}, {Barr}, {Barsotti}, {Barsuglia},
  {Barta}, {Bartlett}, {Barton}, {Bartos}, {Bassiri}, {Basti}, {Batch},
  {Baune}, {Bavigadda}, {Bazzan}, {Behnke}, {Bejger}, {Belczynski}, {Bell},
  {Bell}, {Berger}, {Bergman}, {Bergmann}, {Berry}, {Bersanetti}, {Bertolini},
  {Betzwieser}, {Bhagwat}, {Bhandare}, {Bilenko}, {Billingsley}, {Birch},
  {Birney}, {Birnholtz}, {Biscans}, {Bisht}, {Bitossi}, {Biwer}, {Bizouard},
  {Blackburn}, {Blair}, {Blair}, {Blair}, {Bloemen}, {Bock}, {Bodiya}, {Boer},
  {Bogaert}, {Bogan}, {Bohe}, {Bojtos}, {Bond}, {Bondu}, {Bonnand}, {Boom},
  {Bork}, {Boschi}, {Bose}, {Bouffanais}, {Bozzi}, {Bradaschia}, {Brady},
  {Braginsky}, {Branchesi}, {Brau}, {Briant}, {Brillet}, {Brinkmann},
  {Brisson}, {Brockill}, {Brooks}, {Brown}, {Brown}, {Brown}, {Buchanan},
  {Buikema}, {Bulik}, {Bulten}, {Buonanno}, {Buskulic}, {Buy}, {Byer},
  {Cabero}, {Cadonati}, {Cagnoli}, {Cahillane}, {Bustillo}, {Callister},
  {Calloni}, {Camp}, {Cannon}, {Cao}, {Capano}, {Capocasa}, {Carbognani},
  {Caride}, {Casanueva Diaz}, {Casentini}, {Caudill}, {Cavagli{\`a}},
  {Cavalier}, {Cavalieri}, {Cella}, {Cepeda}, {Baiardi}, {Cerretani},
  {Cesarini}, {Chakraborty}, {Chalermsongsak}, {Chamberlin}, {Chan}, {Chao},
  {Charlton}, {Chassande-Mottin}, {Chen}, {Chen}, {Cheng}, {Chincarini},
  {Chiummo}, {Cho}, {Cho}, {Chow}, {Christensen}, {Chu}, {Chua}, {Chung},
  {Ciani}, {Clara}, {Clark}, {Cleva}, {Coccia}, {Cohadon}, {Colla}, {Collette},
  {Cominsky}, {Constancio}, {Conte}, {Conti}, {Cook}, {Corbitt}, {Cornish},
  {Corsi}, {Cortese}, {Costa}, {Coughlin}, {Coughlin}, {Coulon}, {Countryman},
  {Couvares}, {Cowan}, {Coward}, {Cowart}, {Coyne}, {Coyne}, {Craig},
  {Creighton}, {Creighton}, {Cripe}, {Crowder}, {Cruise}, {Cumming},
  {Cunningham}, {Cuoco}, {Dal Canton}, {Danilishin}, {D'Antonio}, {Danzmann},
  {Darman}, {Da Silva Costa}, {Dattilo}, {Dave}, {Daveloza}, {Davier},
  {Davies}, {Daw}, {Day}, {De}, {DeBra}, {Debreczeni}, {Degallaix}, {De
  Laurentis}, {Del{\'e}glise}, {Del Pozzo}, {Denker}, {Dent}, {Dereli},
  {Dergachev}, {DeRosa}, {De Rosa}, {DeSalvo}, {Dhurandhar}, {D{\'\i}az}, {Di
  Fiore}, {Di Giovanni}, {Di Lieto}, {Di Pace}, {Di Palma}, {Di Virgilio},
  {Dojcinoski}, {Dolique}, {Donovan}, {Dooley}, {Doravari}, {Douglas},
  {Downes}, {Drago}, {Drever}, {Driggers}, {Du}, {Ducrot}, {Dwyer}, {Edo},
  {Edwards}, {Effler}, {Eggenstein}, {Ehrens}, {Eichholz}, {Eikenberry},
  {Engels}, {Essick}, {Etzel}, {Evans}, {Evans}, {Everett}, {Factourovich},
  {Fafone}, {Fair}, {Fairhurst}, {Fan}, {Fang}, {Farinon}, {Farr}, {Farr},
  {Favata}, {Fays}, {Fehrmann}, {Fejer}, {Feldbaum}, {Ferrante}, {Ferreira},
  {Ferrini}, {Fidecaro}, {Finn}, {Fiori}, {Fiorucci}, {Fisher}, {Flaminio},
  {Fletcher}, {Fong}, {Fournier}, {Franco}, {Frasca}, {Frasconi}, {Frede},
  {Frei}, {Freise}, {Frey}, {Frey}, {Fricke}, {Fritschel}, {Frolov}, {Fulda},
  {Fyffe}, {Gabbard}, {Gair}, {Gammaitoni}, {Gaonkar}, {Garufi}, {Gatto},
  {Gaur}, {Gehrels}, {Gemme}, {Gendre}, {Genin}, {Gennai}, {George}, {Gergely},
  {Germain}, {Ghosh}, {Ghosh}, {Ghosh}, {Giaime}, {Giardina}, {Giazotto},
  {Gill}, {Glaefke}, {Gleason}, {Goetz}, {Goetz}, {Gondan}, {Gonz{\'a}lez},
  {Castro}, {Gopakumar}, {Gordon}, {Gorodetsky}, {Gossan}, {Gosselin},
  {Gouaty}, {Graef}, {Graff}, {Granata}, {Grant}, {Gras}, {Gray}, {Greco},
  {Green}, {Greenhalgh}, {Groot}, {Grote}, {Grunewald}, {Guidi}, {Guo},
  {Gupta}, {Gupta}, {Gushwa}, {Gustafson}, {Gustafson}, {Hacker}, {Hall},
  {Hall}, {Hammond}, {Haney}, {Hanke}, {Hanks}, {Hanna}, {Hannam}, {Hanson},
  {Hardwick}, {Harms}, {Harry}, {Harry}, {Hart}, {Hartman}, {Haster},
  {Haughian}, {Healy}, {Heefner}, {Heidmann}, {Heintze}, {Heinzel}, {Heitmann},
  {Hello}, {Hemming}, {Hendry}, {Heng}, {Hennig}, {Heptonstall}, {Heurs},
  {Hild}, {Hoak}, {Hodge}, {Hofman}, {Hollitt}, {Holt}, {Holz}, {Hopkins},
  {Hosken}, {Hough}, {Houston}, {Howell}, {Hu}, {Huang}, {Huerta}, {Huet},
  {Hughey}, {Husa}, {Huttner}, {Huynh-Dinh}, {Idrisy}, {Indik}, {Ingram},
  {Inta}, {Isa}, {Isac}, {Isi}, {Islas}, {Isogai}, {Iyer}, {Izumi}, {Jacobson},
  {Jacqmin}, {Jang}, {Jani}, {Jaranowski}, {Jawahar}, {Jim{\'e}nez-Forteza},
  {Johnson}, {Johnson-McDaniel}, {Jones}, {Jones}, {Jonker}, {Ju}, {Haris},
  {Kalaghatgi}, {Kalogera}, {Kandhasamy}, {Kang}, {Kanner}, {Karki},
  {Kasprzack}, {Katsavounidis}, {Katzman}, {Kaufer}, {Kaur}, {Kawabe},
  {Kawazoe}, {K{\'e}f{\'e}lian}, {Kehl}, {Keitel}, {Kelley}, {Kells},
  {Kennedy}, {Keppel}, {Key}, {Khalaidovski}, {Khalili}, {Khan}, {Khan},
  {Khan}, {Khazanov}, {Kijbunchoo}, {Kim}, {Kim}, {Kim}, {Kim}, {Kim}, {Kim},
  {King}, {King}, {Kinzel}, {Kissel}, {Kleybolte}, {Klimenko}, {Koehlenbeck},
  {Kokeyama}, {Koley}, {Kondrashov}, {Kontos}, {Koranda}, {Korobko}, {Korth},
  {Kowalska}, {Kozak}, {Kringel}, {Krishnan}, {Kr{\'o}lak}, {Krueger}, {Kuehn},
  {Kumar}, {Kumar}, {Kuo}, {Kutynia}, {Kwee}, {Lackey}, {Landry}, {Lange},
  {Lantz}, {Lasky}, {Lazzarini}, {Lazzaro}, {Leaci}, {Leavey}, {Lebigot},
  {Lee}, {Lee}, {Lee}, {Lee}, {Lenon}, {Leonardi}, {Leong}, {Leroy},
  {Letendre}, {Levin}, {Levine}, {Li}, {Libson}, {Littenberg}, {Lockerbie},
  {Logue}, {Lombardi}, {London}, {Lord}, {Lorenzini}, {Loriette}, {Lormand},
  {Losurdo}, {Lough}, {Lousto}, {Lovelace}, {L{\"u}ck}, {Lundgren}, {Luo},
  {Lynch}, {Ma}, {MacDonald}, {Machenschalk}, {MacInnis}, {Macleod},
  {Maga{\~n}a-Sandoval}, {Magee}, {Mageswaran}, {Majorana}, {Maksimovic},
  {Malvezzi}, {Man}, {Mandel}, {Mandic}, {Mangano}, {Mansell}, {Manske},
  {Mantovani}, {Marchesoni}, {Marion}, {M{\'a}rka}, {M{\'a}rka}, {Markosyan},
  {Maros}, {Martelli}, {Martellini}, {Martin}, {Martin}, {Martynov}, {Marx},
  {Mason}, {Masserot}, {Massinger}, {Masso-Reid}, {Matichard}, {Matone},
  {Mavalvala}, {Mazumder}, {Mazzolo}, {McCarthy}, {McClelland}, {McCormick},
  {McGuire}, {McIntyre}, {McIver}, {McManus}, {McWilliams}, {Meacher},
  {Meadors}, {Meidam}, {Melatos}, {Mendell}, {Mendoza-Gandara}, {Mercer},
  {Merilh}, {Merzougui}, {Meshkov}, {Messenger}, {Messick}, {Meyers},
  {Mezzani}, {Miao}, {Michel}, {Middleton}, {Mikhailov}, {Milano}, {Miller},
  {Millhouse}, {Minenkov}, {Ming}, {Mirshekari}, {Mishra}, {Mitra},
  {Mitrofanov}, {Mitselmakher}, {Mittleman}, {Moggi}, {Mohan}, {Mohapatra},
  {Montani}, {Moore}, {Moore}, {Moraru}, {Moreno}, {Morriss}, {Mossavi},
  {Mours}, {Mow-Lowry}, {Mueller}, {Mueller}, {Muir}, {Mukherjee}, {Mukherjee},
  {Mukherjee}, {Mukund}, {Mullavey}, {Munch}, {Murphy}, {Murray}, {Mytidis},
  {Nardecchia}, {Naticchioni}, {Nayak}, {Necula}, {Nedkova}, {Nelemans},
  {Neri}, {Neunzert}, {Newton}, {Nguyen}, {Nielsen}, {Nissanke}, {Nitz},
  {Nocera}, {Nolting}, {Normandin}, {Nuttall}, {Oberling}, {Ochsner}, {O'Dell},
  {Oelker}, {Ogin}, {Oh}, {Oh}, {Ohme}, {Oliver}, {Oppermann}, {Oram},
  {O'Reilly}, {O'Shaughnessy}, {Ott}, {Ottaway}, {Ottens}, {Overmier}, {Owen},
  {Pai}, {Pai}, {Palamos}, {Palashov}, {Palomba}, {Pal-Singh}, {Pan}, {Pan},
  {Pankow}, {Pannarale}, {Pant}, {Paoletti}, {Paoli}, {Papa}, {Paris},
  {Parker}, {Pascucci}, {Pasqualetti}, {Passaquieti}, {Passuello},
  {Patricelli}, {Patrick}, {Pearlstone}, {Pedraza}, {Pedurand}, {Pekowsky},
  {Pele}, {Penn}, {Perreca}, {Pfeiffer}, {Phelps}, {Piccinni}, {Pichot},
  {Pickenpack}, {Piergiovanni}, {Pierro}, {Pillant}, {Pinard}, {Pinto},
  {Pitkin}, {Poeld}, {Poggiani}, {Popolizio}, {Post}, {Powell}, {Prasad},
  {Predoi}, {Premachandra}, {Prestegard}, {Price}, {Prijatelj}, {Principe},
  {Privitera}, {Prix}, {Prodi}, {Prokhorov}, {Puncken}, {Punturo}, {Puppo},
  {P{\"u}rrer}, {Qi}, {Qin}, {Quetschke}, {Quintero}, {Quitzow-James}, {Raab},
  {Rabeling}, {Radkins}, {Raffai}, {Raja}, {Rakhmanov}, {Ramet}, {Rapagnani},
  {Raymond}, {Razzano}, {Re}, {Read}, {Reed}, {Regimbau}, {Rei}, {Reid},
  {Reitze}, {Rew}, {Reyes}, {Ricci}, {Riles}, {Robertson}, {Robie}, {Robinet},
  {Rocchi}, {Rolland}, {Rollins}, {Roma}, {Romano}, {Romano}, {Romanov},
  {Romie}, {Rosi{\'n}ska}, {Rowan}, {R{\"u}diger}, {Ruggi}, {Ryan}, {Sachdev},
  {Sadecki}, {Sadeghian}, {Salconi}, {Saleem}, {Salemi}, {Samajdar}, {Sammut},
  {Sampson}, {Sanchez}, {Sandberg}, {Sandeen}, {Sanders}, {Sanders},
  {Sassolas}, {Sathyaprakash}, {Saulson}, {Sauter}, {Savage}, {Sawadsky},
  {Schale}, {Schilling}, {Schmidt}, {Schmidt}, {Schnabel}, {Schofield},
  {Sch{\"o}nbeck}, {Schreiber}, {Schuette}, {Schutz}, {Scott}, {Scott},
  {Sellers}, {Sengupta}, {Sentenac}, {Sequino}, {Sergeev}, {Serna},
  {Setyawati}, {Sevigny}, {Shaddock}, {Shaffer}, {Shah}, {Shahriar}, {Shaltev},
  {Shao}, {Shapiro}, {Shawhan}, {Sheperd}, {Shoemaker}, {Shoemaker}, {Siellez},
  {Siemens}, {Sigg}, {Silva}, {Simakov}, {Singer}, {Singer}, {Singh}, {Singh},
  {Singhal}, {Sintes}, {Slagmolen}, {Smith}, {Smith}, {Smith}, {Smith}, {Son},
  {Sorazu}, {Sorrentino}, {Souradeep}, {Srivastava}, {Staley}, {Steinke},
  {Steinlechner}, {Steinlechner}, {Steinmeyer}, {Stephens}, {Stevenson},
  {Stone}, {Strain}, {Straniero}, {Stratta}, {Strauss}, {Strigin}, {Sturani},
  {Stuver}, {Summerscales}, {Sun}, {Sutton}, {Swinkels}, {Szczepa{\'n}czyk},
  {Tacca}, {Talukder}, {Tanner}, {T{\'a}pai}, {Tarabrin}, {Taracchini},
  {Taylor}, {Theeg}, {Thirugnanasambandam}, {Thomas}, {Thomas}, {Thomas},
  {Thorne}, {Thorne}, {Thrane}, {Tiwari}, {Tiwari}, {Tokmakov}, {Tomlinson},
  {Tonelli}, {Torres}, {Torrie}, {T{\"o}yr{\"a}}, {Travasso}, {Traylor},
  {Trifir{\`o}}, {Tringali}, {Trozzo}, {Tse}, {Turconi}, {Tuyenbayev},
  {Ugolini}, {Unnikrishnan}, {Urban}, {Usman}, {Vahlbruch}, {Vajente},
  {Valdes}, {Vallisneri}, {van Bakel}, {van Beuzekom}, {van den Brand}, {Van
  Den Broeck}, {Vander-Hyde}, {van der Schaaf}, {van Heijningen}, {van Veggel},
  {Vardaro}, {Vass}, {Vas{\'u}th}, {Vaulin}, {Vecchio}, {Vedovato}, {Veitch},
  {Veitch}, {Venkateswara}, {Verkindt}, {Vetrano}, {Vicer{\'e}}, {Vinciguerra},
  {Vine}, {Vinet}, {Vitale}, {Vo}, {Vocca}, {Vorvick}, {Voss}, {Vousden},
  {Vyatchanin}, {Wade}, {Wade}, {Wade}, {Waldman}, {Walker}, {Wallace},
  {Walsh}, {Wang}, {Wang}, {Wang}, {Wang}, {Wang}, {Ward}, {Ward}, {Warner},
  {Was}, {Weaver}, {Wei}, {Weinert}, {Weinstein}, {Weiss}, {Welborn}, {Wen},
  {We{\ss}els}, {Westphal}, {Wette}, {Whelan}, {Whitcomb}, {White}, {Whiting},
  {Wiesner}, {Wilkinson}, {Willems}, {Williams}, {Williams}, {Williamson},
  {Willis}, {Willke}, {Wimmer}, {Winkelmann}, {Winkler}, {Wipf}, {Wiseman},
  {Wittel}, {Woan}, {Worden}, {Wright}, {Wu}, {Yablon}, {Yakushin}, {Yam},
  {Yamamoto}, {Yancey}, {Yap}, {Yu}, {Yvert}, {Zadro{\.Z}ny}, {Zangrando},
  {Zanolin}, {Zendri}, {Zevin}, {Zhang}, {Zhang}, {Zhang}, {Zhang}, {Zhao},
  {Zhou}, {Zhou}, {Zhu}, {Zucker}, {Zuraw}, {Zweizig}, {LIGO Scientific
  Collaboration}, \& {Virgo Collaboration}}]{2016PhRvL.116f1102A}
{Abbott}, B.~P., {Abbott}, R., {Abbott}, T.~D., {et~al.} 2016,
  \href{http://dx.doi.org/10.1103/PhysRevLett.116.061102}{\color{magenta}\prl},
  \href{https://ui.adsabs.harvard.edu/abs/2016PhRvL.116f1102A}{\color{blue}116},
  \href{https://ui.adsabs.harvard.edu/abs/2016PhRvL.116f1102A}{\color{blue}061102}

\bibitem[{{Abeysekara} {et~al.}(2016){Abeysekara}, {Archambault}, {Archer},
  {Benbow}, {Bird}, {Buchovecky}, {Buckley}, {Byrum}, {Cardenzana}, {Cerruti},
  {Chen}, {Christiansen}, {Ciupik}, {Cui}, {Dickinson}, {Eisch}, {Errando},
  {Falcone}, {Fegan}, {Feng}, {Finley}, {Fleischhack}, {Fortin}, {Fortson},
  {Furniss}, {Gillanders}, {Griffin}, {Grube}, {Gyuk}, {H{\"u}tten},
  {H{\r{a}}kansson}, {Hanna}, {Holder}, {Humensky}, {Johnson}, {Kaaret}, {Kar},
  {Kelley-Hoskins}, {Kertzman}, {Kieda}, {Krause}, {Krennrich}, {Kumar},
  {Lang}, {Lin}, {Maier}, {McArthur}, {McCann}, {Meagher}, {Moriarty},
  {Mukherjee}, {Nieto}, {O'Brien}, {O'Faol{\'a}in de Bhr{\'o}ithe}, {Ong},
  {Otte}, {Park}, {Perkins}, {Petrashyk}, {Pohl}, {Popkow}, {Pueschel},
  {Quinn}, {Ragan}, {Ratliff}, {Reynolds}, {Richards}, {Roache}, {Santander},
  {Sembroski}, {Shahinyan}, {Staszak}, {Telezhinsky}, {Tucci}, {Tyler},
  {Vincent}, {Wakely}, {Weiner}, {Weinstein}, {Williams}, \&
  {Zitzer}}]{2016ApJ...818L..33A}
{Abeysekara}, A.~U., {Archambault}, S., {Archer}, A., {et~al.} 2016,
  \href{http://dx.doi.org/10.3847/2041-8205/818/2/L33}{\color{magenta}\apjl},
  \href{https://ui.adsabs.harvard.edu/abs/2016ApJ...818L..33A}{\color{blue}818},
  \href{https://ui.adsabs.harvard.edu/abs/2016ApJ...818L..33A}{\color{blue}L33}

\bibitem[{{Abeysekara} {et~al.}(2020){Abeysekara}, {Benbow}, {Brill},
  {Buckley}, {Christiansen}, {Chromey}, {Daniel}, {Davis}, {Falcone}, {Feng},
  {Finley}, {Fortson}, {Furniss}, {Gent}, {Giuri}, {Gueta}, {Hanna}, {Hassan},
  {Hervet}, {Holder}, {Hughes}, {Humensky}, {Kaaret}, {Kertzman}, {Kieda},
  {Krennrich}, {Kumar}, {LeBohec}, {Lin}, {Lundy}, {Maier}, {Matthews},
  {Moriarty}, {Mukherjee}, {Nievas-Rosillo}, {O'Brien}, {Ong}, {Otte},
  {Pfrang}, {Pohl}, {Prado}, {Pueschel}, {Quinn}, {Ragan}, {Reynolds},
  {Ribeiro}, {Richards}, {Roache}, {Ryan}, {Santander}, {Sembroski}, {Wakely},
  {Weinstein}, {Wilcox}, {Williams}, \& {Williamson}}]{2020NatAs...4.1164A}
{Abeysekara}, A.~U., {Benbow}, W., {Brill}, A., {et~al.} 2020,
  \href{http://dx.doi.org/10.1038/s41550-020-1143-y}{\color{magenta}NatAs},
  \href{https://ui.adsabs.harvard.edu/abs/2020NatAs...4.1164A}{\color{blue}4},
  \href{https://ui.adsabs.harvard.edu/abs/2020NatAs...4.1164A}{\color{blue}1164}

\bibitem[{{Ajagekar} \& {You}(2020)}]{2020arXiv200300264A}
{Ajagekar}, A., \& {You}, F. 2020,
  \href{https://arxiv.org/abs/2003.00264}{\color{magenta}arXiv},
  \href{https://ui.adsabs.harvard.edu/abs/2020arXiv200300264A}{\color{blue}arXiv:2003.00264}.
\newblock \doarXiv{2003.00264}

\bibitem[{{Arute} {et~al.}(2019){Arute}, {Arya}, {Babbush}, {Bacon}, {Bardin},
  {Barends}, {Biswas}, {Boixo}, {Brandao}, {Buell}, {Burkett}, {Chen}, {Chen},
  {Chiaro}, {Collins}, {Courtney}, {Dunsworth}, {Farhi}, {Foxen}, {Fowler},
  {Gidney}, {Giustina}, {Graff}, {Guerin}, {Habegger}, {Harrigan}, {Hartmann},
  {Ho}, {Hoffmann}, {Huang}, {Humble}, {Isakov}, {Jeffrey}, {Jiang}, {Kafri},
  {Kechedzhi}, {Kelly}, {Klimov}, {Knysh}, {Korotkov}, {Kostritsa}, {Landhuis},
  {Lindmark}, {Lucero}, {Lyakh}, {Mandr{\`a}}, {McClean}, {McEwen}, {Megrant},
  {Mi}, {Michielsen}, {Mohseni}, {Mutus}, {Naaman}, {Neeley}, {Neill}, {Niu},
  {Ostby}, {Petukhov}, {Platt}, {Quintana}, {Rieffel}, {Roushan}, {Rubin},
  {Sank}, {Satzinger}, {Smelyanskiy}, {Sung}, {Trevithick}, {Vainsencher},
  {Villalonga}, {White}, {Yao}, {Yeh}, {Zalcman}, {Neven}, \&
  {Martinis}}]{2019Natur.574..505A}
{Arute}, F., {Arya}, K., {Babbush}, R., {et~al.} 2019,
  \href{http://dx.doi.org/10.1038/s41586-019-1666-5}{\color{magenta}\nat},
  \href{https://ui.adsabs.harvard.edu/abs/2019Natur.574..505A}{\color{blue}574},
  \href{https://ui.adsabs.harvard.edu/abs/2019Natur.574..505A}{\color{blue}505}

\bibitem[{{Banaszek} {et~al.}(2020){Banaszek}, {Kunz}, {Jachura}, \&
  {Jarzyna}}]{2020JLwT...38.2741B}
{Banaszek}, K., {Kunz}, L., {Jachura}, M., \& {Jarzyna}, M. 2020,
  \href{http://dx.doi.org/10.1109/JLT.2020.2973890}{\color{magenta}Journal of
  Lightwave Technology},
  \href{https://ui.adsabs.harvard.edu/abs/2020JLwT...38.2741B}{\color{blue}38},
  \href{https://ui.adsabs.harvard.edu/abs/2020JLwT...38.2741B}{\color{blue}2741}

\bibitem[{{Barstow} \& {Holberg}(2007)}]{2007eua..book.....B}
{Barstow}, M.~A., \& {Holberg}, J.~B. 2007, {Extreme Ultraviolet Astronomy}
  (Cambridge University Press)

\bibitem[{{Bassi} \& {Ghirardi}(2008)}]{2008IJTP...47.2500B}
{Bassi}, A., \& {Ghirardi}, G. 2008,
  \href{http://dx.doi.org/10.1007/s10773-008-9683-4}{\color{magenta}International
  Journal of Theoretical Physics},
  \href{https://ui.adsabs.harvard.edu/abs/2008IJTP...47.2500B}{\color{blue}47},
  \href{https://ui.adsabs.harvard.edu/abs/2008IJTP...47.2500B}{\color{blue}2500}

\bibitem[{{Becerra} {et~al.}(2015){Becerra}, {Fan}, \&
  {Migdall}}]{2015NaPho...9...48B}
{Becerra}, F.~E., {Fan}, J., \& {Migdall}, A. 2015,
  \href{http://dx.doi.org/10.1038/nphoton.2014.280}{\color{magenta}Nature
  Photonics},
  \href{https://ui.adsabs.harvard.edu/abs/2015NaPho...9...48B}{\color{blue}9},
  \href{https://ui.adsabs.harvard.edu/abs/2015NaPho...9...48B}{\color{blue}48}

\bibitem[{{Benedetti} {et~al.}(2016){Benedetti}, {Realpe-G{\'o}mez}, {Biswas},
  \& {Perdomo-Ortiz}}]{2016PhRvA..94b2308B}
{Benedetti}, M., {Realpe-G{\'o}mez}, J., {Biswas}, R., \& {Perdomo-Ortiz}, A.
  2016,
  \href{http://dx.doi.org/10.1103/PhysRevA.94.022308}{\color{magenta}\pra},
  \href{https://ui.adsabs.harvard.edu/abs/2016PhRvA..94b2308B}{\color{blue}94},
  \href{https://ui.adsabs.harvard.edu/abs/2016PhRvA..94b2308B}{\color{blue}022308}

\bibitem[{Bennett {et~al.}(2002)Bennett, Shor, Smolin, \&
  Thapliyal}]{Bennett2002}
Bennett, C., Shor, P., Smolin, J., \& Thapliyal, A. 2002,
  \href{http://dx.doi.org/10.1109/tit.2002.802612}{\color{magenta}{IEEE}
  Transactions on Information Theory}, 48, 2637

\bibitem[{{Bennett} \& {Wiesner}(1992)}]{1992PhRvL..69.2881B}
{Bennett}, C.~H., \& {Wiesner}, S.~J. 1992,
  \href{http://dx.doi.org/10.1103/PhysRevLett.69.2881}{\color{magenta}\prl},
  \href{https://ui.adsabs.harvard.edu/abs/1992PhRvL..69.2881B}{\color{blue}69},
  \href{https://ui.adsabs.harvard.edu/abs/1992PhRvL..69.2881B}{\color{blue}2881}

\bibitem[{{Benton}(2019)}]{2019PASP..131g4501B}
{Benton}, D.~M. 2019,
  \href{http://dx.doi.org/10.1088/1538-3873/ab1a46}{\color{magenta}\pasp},
  \href{https://ui.adsabs.harvard.edu/abs/2019PASP..131g4501B}{\color{blue}131},
  \href{https://ui.adsabs.harvard.edu/abs/2019PASP..131g4501B}{\color{blue}074501}

\bibitem[{{Benton} {et~al.}(2019){Benton}, {Zandi}, \&
  {Sugden}}]{2019SPIE11161E..0GB}
{Benton}, D.~M., {Zandi}, M.~A., \& {Sugden}, K. 2019,
  \href{http://dx.doi.org/10.1117/12.2532058}{\color{magenta}Proc.~SPIE},
  \href{https://ui.adsabs.harvard.edu/abs/2019SPIE11161E..0GB}{\color{blue}11161},
  \href{https://ui.adsabs.harvard.edu/abs/2019SPIE11161E..0GB}{\color{blue}111610G}

\bibitem[{{Berera}(2020)}]{2020PhRvD.102f3005B}
{Berera}, A. 2020,
  \href{http://dx.doi.org/10.1103/PhysRevD.102.063005}{\color{magenta}\prd},
  \href{https://ui.adsabs.harvard.edu/abs/2020PhRvD.102f3005B}{\color{blue}102},
  \href{https://ui.adsabs.harvard.edu/abs/2020PhRvD.102f3005B}{\color{blue}063005}

\bibitem[{{Berkhout} \& {Beijersbergen}(2009)}]{2009JOptA..11i4021B}
{Berkhout}, G.~C.~G., \& {Beijersbergen}, M.~W. 2009,
  \href{http://dx.doi.org/10.1088/1464-4258/11/9/094021}{\color{magenta}Journal
  of Optics A: Pure and Applied Optics},
  \href{https://ui.adsabs.harvard.edu/abs/2009JOptA..11i4021B}{\color{blue}11},
  \href{https://ui.adsabs.harvard.edu/abs/2009JOptA..11i4021B}{\color{blue}094021}

\bibitem[{Beskin {et~al.}(1997)Beskin, Borisov, Komarova, Mitronova,
  Neizvestny, Plokhotnichenko, \& Popova}]{Beskin1997}
Beskin, G., Borisov, N., Komarova, V., {et~al.} 1997,
  \href{http://dx.doi.org/10.1023/a:1000845628229}{\color{magenta}Astrophysics
  and Space Science}, 252, 51

\bibitem[{{Bland-Hawthorn} {et~al.}(2021){Bland-Hawthorn}, {Sellars}, \&
  {Bartholomew}}]{2021arXiv210307590B}
{Bland-Hawthorn}, J., {Sellars}, M., \& {Bartholomew}, J. 2021,
  \href{https://arxiv.org/abs/2103.07590}{\color{magenta}arXiv},
  \href{https://ui.adsabs.harvard.edu/abs/2021arXiv210307590B}{\color{blue}arXiv:2103.07590}.
\newblock \doarXiv{2103.07590}

\bibitem[{{Blasone} {et~al.}(2009){Blasone}, {Dell'Anno}, {DeSiena}, \&
  {Illuminati}}]{2009EL.....8550002B}
{Blasone}, M., {Dell'Anno}, F., {DeSiena}, S., \& {Illuminati}, F. 2009,
  \href{http://dx.doi.org/10.1209/0295-5075/85/50002}{\color{magenta}EPL
  (Europhysics Letters)},
  \href{https://ui.adsabs.harvard.edu/abs/2009EL.....8550002B}{\color{blue}85},
  \href{https://ui.adsabs.harvard.edu/abs/2009EL.....8550002B}{\color{blue}50002}

\bibitem[{{Breitenbach} {et~al.}(1997){Breitenbach}, {Schiller}, \&
  {Mlynek}}]{1997Natur.387..471B}
{Breitenbach}, G., {Schiller}, S., \& {Mlynek}, J. 1997,
  \href{http://dx.doi.org/10.1038/387471a0}{\color{magenta}\nat},
  \href{https://ui.adsabs.harvard.edu/abs/1997Natur.387..471B}{\color{blue}387},
  \href{https://ui.adsabs.harvard.edu/abs/1997Natur.387..471B}{\color{blue}471}

\bibitem[{{Brown} \& {Twiss}(1954)}]{1954PMag...45..663B}
{Brown}, R.~H., \& {Twiss}, R.~G. 1954,
  \href{http://dx.doi.org/10.1080/14786440708520475}{\color{magenta}Philosophical
  Magazine},
  \href{https://ui.adsabs.harvard.edu/abs/1954PMag...45..663B}{\color{blue}45},
  \href{https://ui.adsabs.harvard.edu/abs/1954PMag...45..663B}{\color{blue}663}

\bibitem[{{Caleffi} {et~al.}(2018){Caleffi}, {Cacciapuoti}, \&
  {Bianchi}}]{2018arXiv180504360C}
{Caleffi}, M., {Cacciapuoti}, A.~S., \& {Bianchi}, G. 2018,
  \href{https://arxiv.org/abs/1805.04360}{\color{magenta}arXiv},
  \href{https://ui.adsabs.harvard.edu/abs/2018arXiv180504360C}{\color{blue}arXiv:1805.04360}.
\newblock \doarXiv{1805.04360}

\bibitem[{Cline(2015)}]{cline2015armada}
Cline, E. 2015, Armada (Random House)

\bibitem[{{Cocconi} \& {Morrison}(1959)}]{1959Natur.184..844C}
{Cocconi}, G., \& {Morrison}, P. 1959,
  \href{http://dx.doi.org/10.1038/184844a0}{\color{magenta}\nat},
  \href{http://adsabs.harvard.edu/abs/1959Natur.184..844C}{\color{blue}184},
  \href{http://adsabs.harvard.edu/abs/1959Natur.184..844C}{\color{blue}844}

\bibitem[{{Croke} {et~al.}(2008){Croke}, {Andersson}, \&
  {Barnett}}]{2008PhRvA..77a2113C}
{Croke}, S., {Andersson}, E., \& {Barnett}, S.~M. 2008,
  \href{http://dx.doi.org/10.1103/PhysRevA.77.012113}{\color{magenta}\pra},
  \href{https://ui.adsabs.harvard.edu/abs/2008PhRvA..77a2113C}{\color{blue}77},
  \href{https://ui.adsabs.harvard.edu/abs/2008PhRvA..77a2113C}{\color{blue}012113}

\bibitem[{{da Silva} {et~al.}(2013){da Silva}, {Guha}, \&
  {Dutton}}]{2013PhRvA..87e2320D}
{da Silva}, M.~P., {Guha}, S., \& {Dutton}, Z. 2013,
  \href{http://dx.doi.org/10.1103/PhysRevA.87.052320}{\color{magenta}\pra},
  \href{https://ui.adsabs.harvard.edu/abs/2013PhRvA..87e2320D}{\color{blue}87},
  \href{https://ui.adsabs.harvard.edu/abs/2013PhRvA..87e2320D}{\color{blue}052320}

\bibitem[{{Deng} {et~al.}(2019){Deng}, {Wang}, {Ding}, {Duan}, {Qin}, {Chen},
  {He}, {He}, {Li}, {Li}, {Peng}, {Matekole}, {Byrnes}, {Schneider}, {Kamp},
  {Wang}, {Dowling}, {H{\"o}fling}, {Lu}, {Scully}, \&
  {Pan}}]{2019PhRvL.123h0401D}
{Deng}, Y.-H., {Wang}, H., {Ding}, X., {et~al.} 2019,
  \href{http://dx.doi.org/10.1103/PhysRevLett.123.080401}{\color{magenta}\prl},
  \href{https://ui.adsabs.harvard.edu/abs/2019PhRvL.123h0401D}{\color{blue}123},
  \href{https://ui.adsabs.harvard.edu/abs/2019PhRvL.123h0401D}{\color{blue}080401}

\bibitem[{{Ding} {et~al.}(2017){Ding}, {Pavlichin}, \&
  {Wilde}}]{2017arXiv170508878D}
{Ding}, D., {Pavlichin}, D.~S., \& {Wilde}, M.~M. 2017,
  \href{https://arxiv.org/abs/1705.08878}{\color{magenta}arXiv},
  \href{https://ui.adsabs.harvard.edu/abs/2017arXiv170508878D}{\color{blue}arXiv:1705.08878}.
\newblock \doarXiv{1705.08878}

\bibitem[{Dolinar(1973)}]{dolinar1973optimum}
Dolinar, S.~J. 1973, Research Laboratory of Electronics, MIT, Quarterly
  Progress Report, 11, 115

\bibitem[{{Donohue} {et~al.}(2013){Donohue}, {Agnew}, {Lavoie}, \&
  {Resch}}]{2013PhRvL.111o3602D}
{Donohue}, J.~M., {Agnew}, M., {Lavoie}, J., \& {Resch}, K.~J. 2013,
  \href{http://dx.doi.org/10.1103/PhysRevLett.111.153602}{\color{magenta}\prl},
  \href{https://ui.adsabs.harvard.edu/abs/2013PhRvL.111o3602D}{\color{blue}111},
  \href{https://ui.adsabs.harvard.edu/abs/2013PhRvL.111o3602D}{\color{blue}153602}

\bibitem[{{Eberhard} \& {Ross}(1989)}]{1989FoPhL...2..127E}
{Eberhard}, P.~H., \& {Ross}, R.~R. 1989,
  \href{http://dx.doi.org/10.1007/BF00696109}{\color{magenta}Foundations of
  Physics Letters},
  \href{https://ui.adsabs.harvard.edu/abs/1989FoPhL...2..127E}{\color{blue}2},
  \href{https://ui.adsabs.harvard.edu/abs/1989FoPhL...2..127E}{\color{blue}127}

\bibitem[{{Fanizza} {et~al.}(2020){Fanizza}, {Rosati}, {Skotiniotis},
  {Calsamiglia}, \& {Giovannetti}}]{2020arXiv200606522F}
{Fanizza}, M., {Rosati}, M., {Skotiniotis}, M., {et~al.} 2020,
  \href{https://arxiv.org/abs/2006.06522}{\color{magenta}arXiv},
  \href{https://ui.adsabs.harvard.edu/abs/2020arXiv200606522F}{\color{blue}arXiv:2006.06522}.
\newblock \doarXiv{2006.06522}

\bibitem[{{Fedrizzi} {et~al.}(2009){Fedrizzi}, {Ursin}, {Herbst}, {Nespoli},
  {Prevedel}, {Scheidl}, {Tiefenbacher}, {Jennewein}, \&
  {Zeilinger}}]{2009NatPh...5..389F}
{Fedrizzi}, A., {Ursin}, R., {Herbst}, T., {et~al.} 2009,
  \href{http://dx.doi.org/10.1038/nphys1255}{\color{magenta}Nature Physics},
  \href{https://ui.adsabs.harvard.edu/abs/2009NatPh...5..389F}{\color{blue}5},
  \href{https://ui.adsabs.harvard.edu/abs/2009NatPh...5..389F}{\color{blue}389}

\bibitem[{{Feynman}(1982)}]{1982IJTP...21..467F}
{Feynman}, R.~P. 1982,
  \href{http://dx.doi.org/10.1007/BF02650179}{\color{magenta}International
  Journal of Theoretical Physics},
  \href{https://ui.adsabs.harvard.edu/abs/1982IJTP...21..467F}{\color{blue}21},
  \href{https://ui.adsabs.harvard.edu/abs/1982IJTP...21..467F}{\color{blue}467}

\bibitem[{{Filatov} \& {Auzinsh}(2020)}]{2020arXiv200108867F}
{Filatov}, S., \& {Auzinsh}, M. 2020,
  \href{https://arxiv.org/abs/2001.08867}{\color{magenta}arXiv},
  \href{https://ui.adsabs.harvard.edu/abs/2020arXiv200108867F}{\color{blue}arXiv:2001.08867}.
\newblock \doarXiv{2001.08867}

\bibitem[{{Forgan}(2014)}]{2014JBIS...67..232F}
{Forgan}, D.~H. 2014, Journal of the British Interplanetary Society,
  \href{https://ui.adsabs.harvard.edu/abs/2014JBIS...67..232F}{\color{blue}67},
  \href{https://ui.adsabs.harvard.edu/abs/2014JBIS...67..232F}{\color{blue}232}.
\newblock \doarXiv{1410.7796}

\bibitem[{{Fukugita}(2020)}]{2020IAUGA..30..480F}
{Fukugita}, M. 2020,
  \href{https://ui.adsabs.harvard.edu/abs/2020IAUGA..30..480F}{\color{blue}480}

\bibitem[{{Gagatsos} {et~al.}(2020){Gagatsos}, {Bullock}, \&
  {Bash}}]{2020arXiv200206733G}
{Gagatsos}, C.~N., {Bullock}, M.~S., \& {Bash}, B.~A. 2020,
  \href{https://arxiv.org/abs/2002.06733}{\color{magenta}arXiv},
  \href{https://ui.adsabs.harvard.edu/abs/2020arXiv200206733G}{\color{blue}arXiv:2002.06733}.
\newblock \doarXiv{2002.06733}

\bibitem[{{Gale} {et~al.}(2020){Gale}, {Wandel}, \&
  {Hill}}]{2020IJAsB..19..295G}
{Gale}, J., {Wandel}, A., \& {Hill}, H. 2020,
  \href{http://dx.doi.org/10.1017/S1473550419000260}{\color{magenta}International
  Journal of Astrobiology},
  \href{https://ui.adsabs.harvard.edu/abs/2020IJAsB..19..295G}{\color{blue}19},
  \href{https://ui.adsabs.harvard.edu/abs/2020IJAsB..19..295G}{\color{blue}295}

\bibitem[{{Ghirardi} \& {Romano}(2012)}]{2012JPhA...45w2001G}
{Ghirardi}, G., \& {Romano}, R. 2012,
  \href{http://dx.doi.org/10.1088/1751-8113/45/23/232001}{\color{magenta}Journal
  of Physics A Mathematical General},
  \href{https://ui.adsabs.harvard.edu/abs/2012JPhA...45w2001G}{\color{blue}45},
  \href{https://ui.adsabs.harvard.edu/abs/2012JPhA...45w2001G}{\color{blue}232001}

\bibitem[{{Gingerich}(2003)}]{Gingerich2003}
{Gingerich}, O. 2003, {Interview of Philip Morrison by Owen Gingerich}
  (American Institute of Physics, College Park, MD USA)

\bibitem[{{Giovannetti} {et~al.}(2004){Giovannetti}, {Guha}, {Lloyd},
  {Maccone}, {Shapiro}, \& {Yuen}}]{2004PhRvL..92b7902G}
{Giovannetti}, V., {Guha}, S., {Lloyd}, S., {et~al.} 2004,
  \href{http://dx.doi.org/10.1103/PhysRevLett.92.027902}{\color{magenta}\prl},
  \href{https://ui.adsabs.harvard.edu/abs/2004PhRvL..92b7902G}{\color{blue}92},
  \href{https://ui.adsabs.harvard.edu/abs/2004PhRvL..92b7902G}{\color{blue}027902}

\bibitem[{{Giovannetti} {et~al.}(2006){Giovannetti}, {Lloyd}, \&
  {Maccone}}]{2006PhRvL..96a0401G}
{Giovannetti}, V., {Lloyd}, S., \& {Maccone}, L. 2006,
  \href{http://dx.doi.org/10.1103/PhysRevLett.96.010401}{\color{magenta}\prl},
  \href{https://ui.adsabs.harvard.edu/abs/2006PhRvL..96a0401G}{\color{blue}96},
  \href{https://ui.adsabs.harvard.edu/abs/2006PhRvL..96a0401G}{\color{blue}010401}

\bibitem[{{Glauber}(1963)}]{1963PhRv..131.2766G}
{Glauber}, R.~J. 1963,
  \href{http://dx.doi.org/10.1103/PhysRev.131.2766}{\color{magenta}Physical
  Review},
  \href{https://ui.adsabs.harvard.edu/abs/1963PhRv..131.2766G}{\color{blue}131},
  \href{https://ui.adsabs.harvard.edu/abs/1963PhRv..131.2766G}{\color{blue}2766}

\bibitem[{{Gol'tsman} {et~al.}(2001){Gol'tsman}, {Okunev}, {Chulkova},
  {Lipatov}, {Semenov}, {Smirnov}, {Voronov}, {Dzardanov}, {Williams}, \&
  {Sobolewski}}]{2001ApPhL..79..705G}
{Gol'tsman}, G.~N., {Okunev}, O., {Chulkova}, G., {et~al.} 2001,
  \href{http://dx.doi.org/10.1063/1.1388868}{\color{magenta}Applied Physics
  Letters},
  \href{https://ui.adsabs.harvard.edu/abs/2001ApPhL..79..705G}{\color{blue}79},
  \href{https://ui.adsabs.harvard.edu/abs/2001ApPhL..79..705G}{\color{blue}705}

\bibitem[{Guerrini {et~al.}(2019)Guerrini, Chiani, Win, \&
  Conti}]{Guerrini2019}
Guerrini, S., Chiani, M., Win, M.~Z., \& Conti, A. 2019 ({IEEE})

\bibitem[{{Guha}(2011)}]{2011PhRvL.106x0502G}
{Guha}, S. 2011,
  \href{http://dx.doi.org/10.1103/PhysRevLett.106.240502}{\color{magenta}\prl},
  \href{https://ui.adsabs.harvard.edu/abs/2011PhRvL.106x0502G}{\color{blue}106},
  \href{https://ui.adsabs.harvard.edu/abs/2011PhRvL.106x0502G}{\color{blue}240502}

\bibitem[{{Guha} {et~al.}(2020){Guha}, {Zhuang}, \&
  {Bash}}]{2020arXiv200103934G}
{Guha}, S., {Zhuang}, Q., \& {Bash}, B. 2020,
  \href{https://arxiv.org/abs/2001.03934}{\color{magenta}arXiv},
  \href{https://ui.adsabs.harvard.edu/abs/2020arXiv200103934G}{\color{blue}arXiv:2001.03934}.
\newblock \doarXiv{2001.03934}

\bibitem[{{Hanbury Brown}(1956)}]{1956Natur.178.1046H}
{Hanbury Brown}, R. 1956,
  \href{http://dx.doi.org/10.1038/1781046a0}{\color{magenta}\nat},
  \href{https://ui.adsabs.harvard.edu/abs/1956Natur.178.1046H}{\color{blue}178},
  \href{https://ui.adsabs.harvard.edu/abs/1956Natur.178.1046H}{\color{blue}1046}

\bibitem[{{Hanna} {et~al.}(2009){Hanna}, {Ball}, {Covault}, {Carson},
  {Driscoll}, {Fortin}, {Gingrich}, {Jarvis}, {Kildea}, {Lindner}, {Mueller},
  {Mukherjee}, {Ong}, {Ragan}, {Williams}, \& {Zweerink}}]{2009AsBio...9..345H}
{Hanna}, D.~S., {Ball}, J., {Covault}, C.~E., {et~al.} 2009,
  \href{http://dx.doi.org/10.1089/ast.2008.0256}{\color{magenta}Astrobiology},
  \href{https://ui.adsabs.harvard.edu/abs/2009AsBio...9..345H}{\color{blue}9},
  \href{https://ui.adsabs.harvard.edu/abs/2009AsBio...9..345H}{\color{blue}345}

\bibitem[{{Hao} {et~al.}(2021){Hao}, {Shi}, {Li}, {Zhuang}, \&
  {Zhang}}]{2021arXiv210107482H}
{Hao}, S., {Shi}, H., {Li}, W., {et~al.} 2021,
  \href{https://arxiv.org/abs/2101.07482}{\color{magenta}arXiv},
  \href{https://ui.adsabs.harvard.edu/abs/2021arXiv210107482H}{\color{blue}arXiv:2101.07482}.
\newblock \doarXiv{2101.07482}

\bibitem[{Hausladen {et~al.}(1996)Hausladen, Jozsa, Schumacher, Westmoreland,
  \& Wootters}]{hausladen1996classical}
Hausladen, P., Jozsa, R., Schumacher, B., {et~al.} 1996, Physical Review A, 54,
  1869

\bibitem[{{Helstrom}(1976)}]{Helstrom1976}
{Helstrom}, C.~W. 1976, Quantum Detection and Estimation Theory, Mathematics in
  Science and Engineering (Elsevier Science)

\bibitem[{{Hippke}(2018{\natexlab{a}})}]{2018AcAau.151...53H}
{Hippke}, M. 2018{\natexlab{a}},
  \href{http://dx.doi.org/10.1016/j.actaastro.2018.05.038}{\color{magenta}Acta
  Astronautica},
  \href{https://ui.adsabs.harvard.edu/abs/2018AcAau.151...53H}{\color{blue}151},
  \href{https://ui.adsabs.harvard.edu/abs/2018AcAau.151...53H}{\color{blue}53}

\bibitem[{{Hippke}(2018{\natexlab{b}})}]{2018JApA...39...74H}
---. 2018{\natexlab{b}},
  \href{http://dx.doi.org/10.1007/s12036-018-9565-y}{\color{magenta}Journal of
  Astrophysics and Astronomy},
  \href{https://ui.adsabs.harvard.edu/abs/2018JApA...39...74H}{\color{blue}39},
  \href{https://ui.adsabs.harvard.edu/abs/2018JApA...39...74H}{\color{blue}74}

\bibitem[{{Hippke}(2019)}]{2019PASP..131c4502H}
---. 2019,
  \href{http://dx.doi.org/10.1088/1538-3873/aafbac}{\color{magenta}\pasp},
  \href{https://ui.adsabs.harvard.edu/abs/2019PASP..131c4502H}{\color{blue}131},
  \href{https://ui.adsabs.harvard.edu/abs/2019PASP..131c4502H}{\color{blue}034502}

\bibitem[{{Hippke}(2020)}]{2020AJ....159...85H}
---. 2020,
  \href{http://dx.doi.org/10.3847/1538-3881/ab5dca}{\color{magenta}\aj},
  \href{https://ui.adsabs.harvard.edu/abs/2020AJ....159...85H}{\color{blue}159},
  \href{https://ui.adsabs.harvard.edu/abs/2020AJ....159...85H}{\color{blue}85}

\bibitem[{{Hippke} \& {Forgan}(2017)}]{2017arXiv171105761H}
{Hippke}, M., \& {Forgan}, D.~H. 2017,
  \href{https://arxiv.org/abs/1711.05761}{\color{magenta}arXiv},
  \href{https://ui.adsabs.harvard.edu/abs/2017arXiv171105761H}{\color{blue}arXiv:1711.05761}.
\newblock \doarXiv{1711.05761}

\bibitem[{Holevo(1998)}]{Holevo1998}
Holevo, A. 1998,
  \href{http://dx.doi.org/10.1109/18.651037}{\color{magenta}{IEEE} Transactions
  on Information Theory}, 44, 269

\bibitem[{Holevo(1973)}]{holevo1973bounds}
Holevo, A.~S. 1973, Problemy Peredachi Informatsii, 9, 3

\bibitem[{{Hong} {et~al.}(1987){Hong}, {Ou}, \& {Mandel}}]{1987PhRvL..59.2044H}
{Hong}, C.~K., {Ou}, Z.~Y., \& {Mandel}, L. 1987,
  \href{http://dx.doi.org/10.1103/PhysRevLett.59.2044}{\color{magenta}\prl},
  \href{https://ui.adsabs.harvard.edu/abs/1987PhRvL..59.2044H}{\color{blue}59},
  \href{https://ui.adsabs.harvard.edu/abs/1987PhRvL..59.2044H}{\color{blue}2044}

\bibitem[{{Howard} {et~al.}(2004){Howard}, {Horowitz}, {Wilkinson}, {Coldwell},
  {Groth}, {Jarosik}, {Latham}, {Stefanik}, {Willman}, {Wolff}, \&
  {Zajac}}]{2004ApJ...613.1270H}
{Howard}, A.~W., {Horowitz}, P., {Wilkinson}, D.~T., {et~al.} 2004,
  \href{http://dx.doi.org/10.1086/423300}{\color{magenta}\apj},
  \href{https://ui.adsabs.harvard.edu/abs/2004ApJ...613.1270H}{\color{blue}613},
  \href{https://ui.adsabs.harvard.edu/abs/2004ApJ...613.1270H}{\color{blue}1270}

\bibitem[{Jarzyna {et~al.}(2015)Jarzyna, Kuszaj, \& Banaszek}]{Jarzyna2015}
Jarzyna, M., Kuszaj, P., \& Banaszek, K. 2015,
  \href{http://dx.doi.org/10.1364/oe.23.003170}{\color{magenta}Optics Express},
  23, 3170

\bibitem[{{Johansson} {et~al.}(2013){Johansson}, {Nation}, \&
  {Nori}}]{2013CoPhC.184.1234J}
{Johansson}, J.~R., {Nation}, P.~D., \& {Nori}, F. 2013,
  \href{http://dx.doi.org/10.1016/j.cpc.2012.11.019}{\color{magenta}Computer
  Physics Communications},
  \href{https://ui.adsabs.harvard.edu/abs/2013CoPhC.184.1234J}{\color{blue}184},
  \href{https://ui.adsabs.harvard.edu/abs/2013CoPhC.184.1234J}{\color{blue}1234}

\bibitem[{{Kimble}(2008)}]{2008Natur.453.1023K}
{Kimble}, H.~J. 2008,
  \href{http://dx.doi.org/10.1038/nature07127}{\color{magenta}\nat},
  \href{https://ui.adsabs.harvard.edu/abs/2008Natur.453.1023K}{\color{blue}453},
  \href{https://ui.adsabs.harvard.edu/abs/2008Natur.453.1023K}{\color{blue}1023}

\bibitem[{{Kingsley}(1993)}]{1993SPIE.1867...75K}
{Kingsley}, S.~A. 1993,
  \href{http://dx.doi.org/10.1117/12.150129}{\color{magenta}Proc.~SPIE},
  \href{https://ui.adsabs.harvard.edu/abs/1993SPIE.1867...75K}{\color{blue}1867},
  \href{https://ui.adsabs.harvard.edu/abs/1993SPIE.1867...75K}{\color{blue}75}

\bibitem[{{Kipping} {et~al.}(2020){Kipping}, {Frank}, \&
  {Scharf}}]{2020IJAsB..19..430K}
{Kipping}, D., {Frank}, A., \& {Scharf}, C. 2020,
  \href{http://dx.doi.org/10.1017/S1473550420000208}{\color{magenta}International
  Journal of Astrobiology},
  \href{https://ui.adsabs.harvard.edu/abs/2020IJAsB..19..430K}{\color{blue}19},
  \href{https://ui.adsabs.harvard.edu/abs/2020IJAsB..19..430K}{\color{blue}430}

\bibitem[{{Klimek} {et~al.}(2016){Klimek}, {Jachura}, {Wasilewski}, \&
  {Banaszek}}]{2016JMOp...63.2074K}
{Klimek}, A., {Jachura}, M., {Wasilewski}, W., \& {Banaszek}, K. 2016,
  \href{http://dx.doi.org/10.1080/09500340.2016.1173731}{\color{magenta}Journal
  of Modern Optics},
  \href{https://ui.adsabs.harvard.edu/abs/2016JMOp...63.2074K}{\color{blue}63},
  \href{https://ui.adsabs.harvard.edu/abs/2016JMOp...63.2074K}{\color{blue}2074}

\bibitem[{{K{\"o}nig} \& {Smith}(2013)}]{2013PhRvL.110d0501K}
{K{\"o}nig}, R., \& {Smith}, G. 2013,
  \href{http://dx.doi.org/10.1103/PhysRevLett.110.040501}{\color{magenta}\prl},
  \href{https://ui.adsabs.harvard.edu/abs/2013PhRvL.110d0501K}{\color{blue}110},
  \href{https://ui.adsabs.harvard.edu/abs/2013PhRvL.110d0501K}{\color{blue}040501}

\bibitem[{{Korpela} {et~al.}(2011){Korpela}, {Anderson}, {Bankay}, {Cobb},
  {Howard}, {Lebofsky}, {Siemion}, {von Korff}, \&
  {Werthimer}}]{2011SPIE.8152E..12K}
{Korpela}, E.~J., {Anderson}, D.~P., {Bankay}, R., {et~al.} 2011,
  \href{http://dx.doi.org/10.1117/12.894066}{\color{magenta}Proc.~SPIE},
  \href{https://ui.adsabs.harvard.edu/abs/2011SPIE.8152E..12K}{\color{blue}8152},
  \href{https://ui.adsabs.harvard.edu/abs/2011SPIE.8152E..12K}{\color{blue}815212}

\bibitem[{{Korzh} {et~al.}(2020){Korzh}, {Zhao}, {Allmaras}, {Frasca}, {Autry},
  {Bersin}, {Beyer}, {Briggs}, {Bumble}, {Colangelo}, {Crouch}, {Dane},
  {Gerrits}, {Lita}, {Marsili}, {Moody}, {Pe{\~n}a}, {Ramirez}, {Rezac},
  {Sinclair}, {Stevens}, {Velasco}, {Verma}, {Wollman}, {Xie}, {Zhu}, {Hale},
  {Spiropulu}, {Silverman}, {Mirin}, {Nam}, {Kozorezov}, {Shaw}, \&
  {Berggren}}]{2020NaPho..14..250K}
{Korzh}, B., {Zhao}, Q.-Y., {Allmaras}, J.~P., {et~al.} 2020,
  \href{http://dx.doi.org/10.1038/s41566-020-0589-x}{\color{magenta}Nature
  Photonics},
  \href{https://ui.adsabs.harvard.edu/abs/2020NaPho..14..250K}{\color{blue}14},
  \href{https://ui.adsabs.harvard.edu/abs/2020NaPho..14..250K}{\color{blue}250}

\bibitem[{{Kumar} {et~al.}(2019){Kumar}, {Lauk}, \&
  {Simon}}]{2019QS&T....4d5003K}
{Kumar}, S., {Lauk}, N., \& {Simon}, C. 2019,
  \href{http://dx.doi.org/10.1088/2058-9565/ab2c87}{\color{magenta}Quantum
  Science and Technology},
  \href{https://ui.adsabs.harvard.edu/abs/2019QS&T....4d5003K}{\color{blue}4},
  \href{https://ui.adsabs.harvard.edu/abs/2019QS&T....4d5003K}{\color{blue}045003}

\bibitem[{{Lacki} {et~al.}(2020){Lacki}, {Brzycki}, {Croft}, {Czech}, {DeBoer},
  {DeMarines}, {Gajjar}, {Isaacson}, {Lebofsky}, {MacMahon}, {Price}, {Sheikh},
  {Siemion}, {Drew}, \& {Worden}}]{2020arXiv200611304L}
{Lacki}, B.~C., {Brzycki}, B., {Croft}, S., {et~al.} 2020,
  \href{https://arxiv.org/abs/2006.11304}{\color{magenta}arXiv},
  \href{https://ui.adsabs.harvard.edu/abs/2020arXiv200611304L}{\color{blue}arXiv:2006.11304}.
\newblock \doarXiv{2006.11304}

\bibitem[{{Leeb} {et~al.}(2013){Leeb}, {Poppe}, {Hammel}, {Alves}, {Brunner},
  \& {Meingast}}]{2013AsBio..13..521L}
{Leeb}, W.~R., {Poppe}, A., {Hammel}, E., {et~al.} 2013,
  \href{http://dx.doi.org/10.1089/ast.2012.0951}{\color{magenta}Astrobiology},
  \href{https://ui.adsabs.harvard.edu/abs/2013AsBio..13..521L}{\color{blue}13},
  \href{https://ui.adsabs.harvard.edu/abs/2013AsBio..13..521L}{\color{blue}521}

\bibitem[{{Li} {et~al.}(2020){Li}, {Wang}, \& {Wang}}]{2020MPLB...3450017L}
{Li}, L.-Y., {Wang}, T.-J., \& {Wang}, C. 2020,
  \href{http://dx.doi.org/10.1142/S0217984920500177}{\color{magenta}Modern
  Physics Letters B},
  \href{https://ui.adsabs.harvard.edu/abs/2020MPLB...3450017L}{\color{blue}34},
  \href{https://ui.adsabs.harvard.edu/abs/2020MPLB...3450017L}{\color{blue}2050017}

\bibitem[{{Lloyd}(1996)}]{1996Sci...273.1073L}
{Lloyd}, S. 1996,
  \href{http://dx.doi.org/10.1126/science.273.5278.1073}{\color{magenta}Science},
  \href{https://ui.adsabs.harvard.edu/abs/1996Sci...273.1073L}{\color{blue}273},
  \href{https://ui.adsabs.harvard.edu/abs/1996Sci...273.1073L}{\color{blue}1073}

\bibitem[{{Lubin}(2016{\natexlab{a}})}]{2016JBIS...69...40L}
{Lubin}, P. 2016{\natexlab{a}}, Journal of the British Interplanetary Society,
  \href{https://ui.adsabs.harvard.edu/abs/2016JBIS...69...40L}{\color{blue}69},
  \href{https://ui.adsabs.harvard.edu/abs/2016JBIS...69...40L}{\color{blue}40}.
\newblock \doarXiv{1604.01356}

\bibitem[{{Lubin}(2016{\natexlab{b}})}]{2016SPIE.9981E..0HL}
{Lubin}, P. 2016{\natexlab{b}},
  \href{http://dx.doi.org/10.1117/12.2238212}{\color{magenta}Proc.~SPIE},
  \href{https://ui.adsabs.harvard.edu/abs/2016SPIE.9981E..0HL}{\color{blue}9981},
  \href{https://ui.adsabs.harvard.edu/abs/2016SPIE.9981E..0HL}{\color{blue}99810H}

\bibitem[{{Maiman}(1960)}]{1960Natur.187..493M}
{Maiman}, T.~H. 1960,
  \href{http://dx.doi.org/10.1038/187493a0}{\color{magenta}\nat},
  \href{http://adsabs.harvard.edu/abs/1960Natur.187..493M}{\color{blue}187},
  \href{http://adsabs.harvard.edu/abs/1960Natur.187..493M}{\color{blue}493}

\bibitem[{{Maire} {et~al.}(2014){Maire}, {Wright}, {Werthimer}, {Treffers},
  {Marcy}, {Stone}, {Drake}, \& {Siemion}}]{2014SPIE.9147E..4KM}
{Maire}, J., {Wright}, S.~A., {Werthimer}, D., {et~al.} 2014,
  \href{http://dx.doi.org/10.1117/12.2056372}{\color{magenta}Proc.~SPIE},
  \href{https://ui.adsabs.harvard.edu/abs/2014SPIE.9147E..4KM}{\color{blue}9147},
  \href{https://ui.adsabs.harvard.edu/abs/2014SPIE.9147E..4KM}{\color{blue}91474K}

\bibitem[{{Maire} {et~al.}(2019){Maire}, {Wright}, {Barrett}, {Dexter},
  {Dorval}, {Duenas}, {Drake}, {Hultgren}, {Isaacson}, {Marcy}, {Meyer},
  {Ramos}, {Shirman}, {Siemion}, {Stone}, {Tallis}, {Tellis}, {Treffers}, \&
  {Werthimer}}]{2019AJ....158..203M}
{Maire}, J., {Wright}, S.~A., {Barrett}, C.~T., {et~al.} 2019,
  \href{http://dx.doi.org/10.3847/1538-3881/ab44d3}{\color{magenta}\aj},
  \href{https://ui.adsabs.harvard.edu/abs/2019AJ....158..203M}{\color{blue}158},
  \href{https://ui.adsabs.harvard.edu/abs/2019AJ....158..203M}{\color{blue}203}

\bibitem[{{Mann} \& {Kimura}(2000)}]{2000JGR...10510317M}
{Mann}, I., \& {Kimura}, H. 2000,
  \href{http://dx.doi.org/10.1029/1999JA900404}{\color{magenta}JGR},
  \href{https://ui.adsabs.harvard.edu/abs/2000JGR...10510317M}{\color{blue}105},
  \href{https://ui.adsabs.harvard.edu/abs/2000JGR...10510317M}{\color{blue}10317}

\bibitem[{Manzalini(2020)}]{quantum2010014}
Manzalini, A. 2020,
  \href{http://dx.doi.org/10.3390/quantum2010014}{\color{magenta}Quantum
  Reports}, 2, 221

\bibitem[{{Marcy}(2021)}]{2021arXiv210201910M}
{Marcy}, G.~W. 2021,
  \href{https://arxiv.org/abs/2102.01910}{\color{magenta}arXiv},
  \href{https://ui.adsabs.harvard.edu/abs/2021arXiv210201910M}{\color{blue}arXiv:2102.01910}.
\newblock \doarXiv{2102.01910}

\bibitem[{{Matekole} {et~al.}(2020){Matekole}, {Cuozzo}, {Prajapati}, {Bhusal},
  {Lee}, {Novikova}, {Mikhailov}, {Dowling}, \& {Cohen}}]{2020PhRvL.125k3602M}
{Matekole}, E.~S., {Cuozzo}, S.~L., {Prajapati}, N., {et~al.} 2020,
  \href{http://dx.doi.org/10.1103/PhysRevLett.125.113602}{\color{magenta}\prl},
  \href{https://ui.adsabs.harvard.edu/abs/2020PhRvL.125k3602M}{\color{blue}125},
  \href{https://ui.adsabs.harvard.edu/abs/2020PhRvL.125k3602M}{\color{blue}113602}

\bibitem[{{Mead}(2013)}]{2013PhDT.......161M}
{Mead}, C.~C. 2013, PhD thesis, Harvard University

\bibitem[{{Merali}(2016)}]{2016Sci...352.1040M}
{Merali}, Z. 2016, Science,
  \href{https://ui.adsabs.harvard.edu/abs/2016Sci...352.1040M}{\color{blue}352},
  \href{https://ui.adsabs.harvard.edu/abs/2016Sci...352.1040M}{\color{blue}1040}

\bibitem[{{M{\"u}ller} {et~al.}(2018){M{\"u}ller}, {Seshadreesan},
  {Peuntinger}, \& {Marquardt}}]{2018arXiv181109423M}
{M{\"u}ller}, C.~R., {Seshadreesan}, K.~P., {Peuntinger}, C., \& {Marquardt},
  C. 2018, \href{https://arxiv.org/abs/1811.09423}{\color{magenta}arXiv},
  \href{https://ui.adsabs.harvard.edu/abs/2018arXiv181109423M}{\color{blue}arXiv:1811.09423}.
\newblock \doarXiv{1811.09423}

\bibitem[{{Oesch} \& {Sanchez}(2014)}]{2014A&A...567A.114O}
{Oesch}, D.~W., \& {Sanchez}, D.~J. 2014,
  \href{http://dx.doi.org/10.1051/0004-6361/201323140}{\color{magenta}\aap},
  \href{https://ui.adsabs.harvard.edu/abs/2014A&A...567A.114O}{\color{blue}567},
  \href{https://ui.adsabs.harvard.edu/abs/2014A&A...567A.114O}{\color{blue}A114}

\bibitem[{{Oi} {et~al.}(2013){Oi}, {Poto{\v{c}}ek}, \&
  {Jeffers}}]{2013PhRvL.110u0504O}
{Oi}, D. K.~L., {Poto{\v{c}}ek}, V., \& {Jeffers}, J. 2013,
  \href{http://dx.doi.org/10.1103/PhysRevLett.110.210504}{\color{magenta}\prl},
  \href{https://ui.adsabs.harvard.edu/abs/2013PhRvL.110u0504O}{\color{blue}110},
  \href{https://ui.adsabs.harvard.edu/abs/2013PhRvL.110u0504O}{\color{blue}210504}

\bibitem[{{Padgett}(2017)}]{2017OExpr..2511265P}
{Padgett}, M.~J. 2017,
  \href{http://dx.doi.org/10.1364/OE.25.011265}{\color{magenta}Optics Express},
  \href{https://ui.adsabs.harvard.edu/abs/2017OExpr..2511265P}{\color{blue}25},
  \href{https://ui.adsabs.harvard.edu/abs/2017OExpr..2511265P}{\color{blue}11265}

\bibitem[{{Padgett} {et~al.}(2015){Padgett}, {Miatto}, {Lavery}, {Zeilinger},
  \& {Boyd}}]{2015NJPh...17b3011P}
{Padgett}, M.~J., {Miatto}, F.~M., {Lavery}, M. P.~J., {et~al.} 2015,
  \href{http://dx.doi.org/10.1088/1367-2630/17/2/023011}{\color{magenta}New
  Journal of Physics},
  \href{https://ui.adsabs.harvard.edu/abs/2015NJPh...17b3011P}{\color{blue}17},
  \href{https://ui.adsabs.harvard.edu/abs/2015NJPh...17b3011P}{\color{blue}023011}

\bibitem[{{Polyakov}(2013)}]{2013ExMPS..45...69P}
{Polyakov}, S.~V. 2013,
  \href{http://dx.doi.org/10.1016/B978-0-12-387695-9.00003-2}{\color{magenta}Single-Photon
  Generation and Detection - Physics and Applications. Series: Experimental
  Methods in the Physical Sciences},
  \href{https://ui.adsabs.harvard.edu/abs/2013ExMPS..45...69P}{\color{blue}45},
  \href{https://ui.adsabs.harvard.edu/abs/2013ExMPS..45...69P}{\color{blue}69}

\bibitem[{{Popkin}(2017)}]{2017Natur.542...20P}
{Popkin}, G. 2017,
  \href{http://dx.doi.org/10.1038/542020a}{\color{magenta}\nat},
  \href{https://ui.adsabs.harvard.edu/abs/2017Natur.542...20P}{\color{blue}542},
  \href{https://ui.adsabs.harvard.edu/abs/2017Natur.542...20P}{\color{blue}20}

\bibitem[{{Price} {et~al.}(2020){Price}, {Enriquez}, {Brzycki}, {Croft},
  {Czech}, {DeBoer}, {DeMarines}, {Foster}, {Gajjar}, {Gizani}, {Hellbourg},
  {Isaacson}, {Lacki}, {Lebofsky}, {MacMahon}, {Pater}, {Siemion}, {Werthimer},
  {Green}, {Kaczmarek}, {Maddalena}, {Mader}, {Drew}, \&
  {Worden}}]{2020AJ....159...86P}
{Price}, D.~C., {Enriquez}, J.~E., {Brzycki}, B., {et~al.} 2020,
  \href{http://dx.doi.org/10.3847/1538-3881/ab65f1}{\color{magenta}\aj},
  \href{https://ui.adsabs.harvard.edu/abs/2020AJ....159...86P}{\color{blue}159},
  \href{https://ui.adsabs.harvard.edu/abs/2020AJ....159...86P}{\color{blue}86}

\bibitem[{{Prochazka} {et~al.}(2013){Prochazka}, {Kodet}, \&
  {Blazej}}]{2013RScI...84d6107P}
{Prochazka}, I., {Kodet}, J., \& {Blazej}, J. 2013,
  \href{http://dx.doi.org/10.1063/1.4802950}{\color{magenta}Review of
  Scientific Instruments},
  \href{https://ui.adsabs.harvard.edu/abs/2013RScI...84d6107P}{\color{blue}84},
  \href{https://ui.adsabs.harvard.edu/abs/2013RScI...84d6107P}{\color{blue}046107}

\bibitem[{{Reines} \& {Marcy}(2002)}]{2002PASP..114..416R}
{Reines}, A.~E., \& {Marcy}, G.~W. 2002,
  \href{http://dx.doi.org/10.1086/342496}{\color{magenta}\pasp},
  \href{https://ui.adsabs.harvard.edu/abs/2002PASP..114..416R}{\color{blue}114},
  \href{https://ui.adsabs.harvard.edu/abs/2002PASP..114..416R}{\color{blue}416}

\bibitem[{{Saglamyurek} {et~al.}(2011){Saglamyurek}, {Sinclair}, {Jin},
  {Slater}, {Oblak}, {Bussi{\`e}res}, {George}, {Ricken}, {Sohler}, \&
  {Tittel}}]{2011Natur.469..512S}
{Saglamyurek}, E., {Sinclair}, N., {Jin}, J., {et~al.} 2011,
  \href{http://dx.doi.org/10.1038/nature09719}{\color{magenta}\nat},
  \href{https://ui.adsabs.harvard.edu/abs/2011Natur.469..512S}{\color{blue}469},
  \href{https://ui.adsabs.harvard.edu/abs/2011Natur.469..512S}{\color{blue}512}

\bibitem[{{Saha}(2019)}]{2019MNRAS.486.5400S}
{Saha}, P. 2019,
  \href{http://dx.doi.org/10.1093/mnras/stz1208}{\color{magenta}\mnras},
  \href{https://ui.adsabs.harvard.edu/abs/2019MNRAS.486.5400S}{\color{blue}486},
  \href{https://ui.adsabs.harvard.edu/abs/2019MNRAS.486.5400S}{\color{blue}5400}

\bibitem[{Saleh {et~al.}(1991)Saleh, Teich, \& John Wiley
  \&~Sons}]{saleh1991fundamentals}
Saleh, B., Teich, M., \& John Wiley \&~Sons, L. 1991, Fundamentals of
  Photonics, Wiley Series in Pure and Applied Optics (Wiley)

\bibitem[{{Santana} {et~al.}(2017){Santana}, {Ma}, {Malein}, {Bastiman},
  {Clarke}, \& {Gerardot}}]{2017PhRvB..95t1410S}
{Santana}, T.~S., {Ma}, Y., {Malein}, R. N.~E., {et~al.} 2017,
  \href{http://dx.doi.org/10.1103/PhysRevB.95.201410}{\color{magenta}\prb},
  \href{https://ui.adsabs.harvard.edu/abs/2017PhRvB..95t1410S}{\color{blue}95},
  \href{https://ui.adsabs.harvard.edu/abs/2017PhRvB..95t1410S}{\color{blue}201410}

\bibitem[{{Schnabel}(2017)}]{2017PhR...684....1S}
{Schnabel}, R. 2017,
  \href{http://dx.doi.org/10.1016/j.physrep.2017.04.001}{\color{magenta}\physrep},
  \href{https://ui.adsabs.harvard.edu/abs/2017PhR...684....1S}{\color{blue}684},
  \href{https://ui.adsabs.harvard.edu/abs/2017PhR...684....1S}{\color{blue}1}

\bibitem[{{Schuetz} {et~al.}(2016){Schuetz}, {Vakoch}, {Shostak}, \&
  {Richards}}]{2016ApJ...825L...5S}
{Schuetz}, M., {Vakoch}, D.~A., {Shostak}, S., \& {Richards}, J. 2016,
  \href{http://dx.doi.org/10.3847/2041-8205/825/1/L5}{\color{magenta}\apjl},
  \href{https://ui.adsabs.harvard.edu/abs/2016ApJ...825L...5S}{\color{blue}825},
  \href{https://ui.adsabs.harvard.edu/abs/2016ApJ...825L...5S}{\color{blue}L5}

\bibitem[{{Schumacher} \& {Westmoreland}(1997)}]{1997PhRvA..56..131S}
{Schumacher}, B., \& {Westmoreland}, M.~D. 1997,
  \href{http://dx.doi.org/10.1103/PhysRevA.56.131}{\color{magenta}\pra},
  \href{https://ui.adsabs.harvard.edu/abs/1997PhRvA..56..131S}{\color{blue}56},
  \href{https://ui.adsabs.harvard.edu/abs/1997PhRvA..56..131S}{\color{blue}131}

\bibitem[{{Schwartz} \& {Townes}(1961)}]{1961Natur.190..205S}
{Schwartz}, R.~N., \& {Townes}, C.~H. 1961,
  \href{http://dx.doi.org/10.1038/190205a0}{\color{magenta}\nat},
  \href{http://adsabs.harvard.edu/abs/1961Natur.190..205S}{\color{blue}190},
  \href{http://adsabs.harvard.edu/abs/1961Natur.190..205S}{\color{blue}205}

\bibitem[{{Sheikh}(2020)}]{2020IJAsB..19..237S}
{Sheikh}, S.~Z. 2020,
  \href{http://dx.doi.org/10.1017/S1473550419000284}{\color{magenta}International
  Journal of Astrobiology},
  \href{https://ui.adsabs.harvard.edu/abs/2020IJAsB..19..237S}{\color{blue}19},
  \href{https://ui.adsabs.harvard.edu/abs/2020IJAsB..19..237S}{\color{blue}237}

\bibitem[{{Sheikh} {et~al.}(2020){Sheikh}, {Siemion}, {Enriquez}, {Price},
  {Isaacson}, {Lebofsky}, {Gajjar}, \& {Kalas}}]{2020AJ....160...29S}
{Sheikh}, S.~Z., {Siemion}, A., {Enriquez}, J.~E., {et~al.} 2020,
  \href{http://dx.doi.org/10.3847/1538-3881/ab9361}{\color{magenta}\aj},
  \href{https://ui.adsabs.harvard.edu/abs/2020AJ....160...29S}{\color{blue}160},
  \href{https://ui.adsabs.harvard.edu/abs/2020AJ....160...29S}{\color{blue}29}

\bibitem[{{Shvartsman}(1981)}]{1981psec.book..122S}
{Shvartsman}, V.~F. 1981, {The MANIIa experiment and potentialities of the
  optical search for extraterrestrial civilizations} ({Goldberg}, H.~S. and
  {Scadron}, M.~D.),
  \href{http://adsabs.harvard.edu/abs/1981psec.book..122S}{\color{blue}122}

\bibitem[{{Stancil} {et~al.}(2012){Stancil}, {Adamson}, {Alania}, {Aliaga},
  {Andrews}, {Del Castillo}, {Bagby}, {Bazo Alba}, {Bodek}, {Boehnlein},
  {Bradford}, {Brooks}, {Budd}, {Butkevich}, {Caicedo}, {Capista},
  {Castromonte}, {Chamorro}, {Charlton}, {Christy}, {Chvojka}, {Conrow},
  {Danko}, {Day}, {Devan}, {Downey}, {Dytman}, {Eberly}, {Fein}, {Felix},
  {Fields}, {Fiorentini}, {Gago}, {Gallagher}, {Gran}, {Grange}, {Griffin},
  {Griffin}, {Hahn}, {Harris}, {Higuera}, {Hobbs}, {Hoffman}, {Hughes},
  {Hurtado}, {Judd}, {Kafka}, {Kephart}, {Kilmer}, {Kordosky}, {Kulagin},
  {Kuznetsov}, {Lanari}, {Le}, {Lee}, {Loiacono}, {Maggi}, {Maher}, {Manly},
  {Mann}, {Marshall}, {McFarland}, {Mislivec}, {McGowan}, {Morf{\'\i}n}, {da
  Motta}, {Mousseau}, {Nelson}, {Niemiec-Gielata}, {Ochoa}, {Osmanov}, {Osta},
  {Palomino}, {Paradis}, {Paolone}, {Park}, {Pe{\~n}a}, {Perdue}, {P{\'e}rez
  Lara}, {Peterman}, {Pla-Dalmau}, {Pollock}, {Prokoshin}, {Ransome}, {Ray},
  {Reyhan}, {Rubinov}, {Ruggiero}, {Sands}, {Schellman}, {Schmitz}, {Schulte},
  {Simon}, {Solano Salinas}, {Stefanski}, {Stevens}, {Tagg}, {Takhistov},
  {Tice}, {Tilden}, {Vel{\'a}squez}, {Vergalosova}, {Voirin}, {Walding},
  {Walker}, {Walton}, {Wolcott}, {Wytock}, {Zavala}, {Zhang}, {Zhu}, \&
  {Ziemer}}]{2012MPLA...2750077S}
{Stancil}, D.~D., {Adamson}, P., {Alania}, M., {et~al.} 2012,
  \href{http://dx.doi.org/10.1142/S0217732312500770}{\color{magenta}Modern
  Physics Letters A},
  \href{https://ui.adsabs.harvard.edu/abs/2012MPLA...2750077S}{\color{blue}27},
  \href{https://ui.adsabs.harvard.edu/abs/2012MPLA...2750077S}{\color{blue}1250077}

\bibitem[{{Stanton}(2019)}]{2019AcAau.156...92S}
{Stanton}, R.~H. 2019,
  \href{http://dx.doi.org/10.1016/j.actaastro.2018.05.061}{\color{magenta}Acta
  Astronautica},
  \href{https://ui.adsabs.harvard.edu/abs/2019AcAau.156...92S}{\color{blue}156},
  \href{https://ui.adsabs.harvard.edu/abs/2019AcAau.156...92S}{\color{blue}92}

\bibitem[{{Subotowicz}(1979)}]{1979AcAau...6..213S}
{Subotowicz}, M. 1979,
  \href{http://dx.doi.org/10.1016/0094-5765(79)90157-7}{\color{magenta}Acta
  Astronautica},
  \href{https://ui.adsabs.harvard.edu/abs/1979AcAau...6..213S}{\color{blue}6},
  \href{https://ui.adsabs.harvard.edu/abs/1979AcAau...6..213S}{\color{blue}213}

\bibitem[{{Tamburini} {et~al.}(2020){Tamburini}, {Thid{\'e}}, \& {Della
  Valle}}]{2020MNRAS.492L..22T}
{Tamburini}, F., {Thid{\'e}}, B., \& {Della Valle}, M. 2020,
  \href{http://dx.doi.org/10.1093/mnrasl/slz176}{\color{magenta}\mnras},
  \href{https://ui.adsabs.harvard.edu/abs/2020MNRAS.492L..22T}{\color{blue}492},
  \href{https://ui.adsabs.harvard.edu/abs/2020MNRAS.492L..22T}{\color{blue}L22}

\bibitem[{{Tan} \& {Kurtsiefer}(2017)}]{2017MNRAS.469.1617T}
{Tan}, P.~K., \& {Kurtsiefer}, C. 2017,
  \href{http://dx.doi.org/10.1093/mnras/stx968}{\color{magenta}\mnras},
  \href{https://ui.adsabs.harvard.edu/abs/2017MNRAS.469.1617T}{\color{blue}469},
  \href{https://ui.adsabs.harvard.edu/abs/2017MNRAS.469.1617T}{\color{blue}1617}

\bibitem[{{Tan} {et~al.}(2014){Tan}, {Yeo}, {Poh}, {Chan}, \&
  {Kurtsiefer}}]{2014ApJ...789L..10T}
{Tan}, P.~K., {Yeo}, G.~H., {Poh}, H.~S., {et~al.} 2014,
  \href{http://dx.doi.org/10.1088/2041-8205/789/1/L10}{\color{magenta}\apjl},
  \href{https://ui.adsabs.harvard.edu/abs/2014ApJ...789L..10T}{\color{blue}789},
  \href{https://ui.adsabs.harvard.edu/abs/2014ApJ...789L..10T}{\color{blue}L10}

\bibitem[{{Tellis} \& {Marcy}(2017)}]{2017AJ....153..251T}
{Tellis}, N.~K., \& {Marcy}, G.~W. 2017,
  \href{http://dx.doi.org/10.3847/1538-3881/aa6d12}{\color{magenta}\aj},
  \href{https://ui.adsabs.harvard.edu/abs/2017AJ....153..251T}{\color{blue}153},
  \href{https://ui.adsabs.harvard.edu/abs/2017AJ....153..251T}{\color{blue}251}

\bibitem[{{Torner} {et~al.}(2005){Torner}, {Torres}, \&
  {Carrasco}}]{2005OExpr..13..873T}
{Torner}, L., {Torres}, J.~P., \& {Carrasco}, S. 2005,
  \href{http://dx.doi.org/10.1364/OPEX.13.000873}{\color{magenta}Optics
  Express},
  \href{https://ui.adsabs.harvard.edu/abs/2005OExpr..13..873T}{\color{blue}13},
  \href{https://ui.adsabs.harvard.edu/abs/2005OExpr..13..873T}{\color{blue}873}

\bibitem[{{Tosi} {et~al.}(2009){Tosi}, {Mora}, {Zappa}, \&
  {Cova}}]{2009JMOp...56..299T}
{Tosi}, A., {Mora}, A.~D., {Zappa}, F., \& {Cova}, S. 2009,
  \href{http://dx.doi.org/10.1080/09500340802263075}{\color{magenta}Journal of
  Modern Optics},
  \href{https://ui.adsabs.harvard.edu/abs/2009JMOp...56..299T}{\color{blue}56},
  \href{https://ui.adsabs.harvard.edu/abs/2009JMOp...56..299T}{\color{blue}299}

\bibitem[{{Uribe-Patarroyo} {et~al.}(2011){Uribe-Patarroyo}, {Alvarez-Herrero},
  {L{\'o}pez Ariste}, {Asensio Ramos}, {Belenguer}, {Manso Sainz}, {Lemen}, \&
  {Gelly}}]{2011A&A...526A..56U}
{Uribe-Patarroyo}, N., {Alvarez-Herrero}, A., {L{\'o}pez Ariste}, A., {et~al.}
  2011,
  \href{http://dx.doi.org/10.1051/0004-6361/201014844}{\color{magenta}\aap},
  \href{https://ui.adsabs.harvard.edu/abs/2011A&A...526A..56U}{\color{blue}526},
  \href{https://ui.adsabs.harvard.edu/abs/2011A&A...526A..56U}{\color{blue}A56}

\bibitem[{Vogel \& Welsch(2006)}]{vogel2006quantum}
Vogel, W., \& Welsch, D. 2006, Quantum Optics (Wiley)

\bibitem[{Vourdas \& da~Rocha(1994)}]{Vourdas1994}
Vourdas, A., \& da~Rocha, J. 1994,
  \href{http://dx.doi.org/10.1080/09500349414552141}{\color{magenta}Journal of
  Modern Optics}, 41, 2291

\bibitem[{{Wang} \& {Fan}(2012)}]{2012JOSAB..29...15W}
{Wang}, S., \& {Fan}, H.-Y. 2012,
  \href{http://dx.doi.org/10.1364/JOSAB.29.000015}{\color{magenta}Journal of
  the Optical Society of America B Optical Physics},
  \href{https://ui.adsabs.harvard.edu/abs/2012JOSAB..29...15W}{\color{blue}29},
  \href{https://ui.adsabs.harvard.edu/abs/2012JOSAB..29...15W}{\color{blue}15}

\bibitem[{{Wilde} {et~al.}(2012){Wilde}, {Guha}, {Tan}, \&
  {Lloyd}}]{2012arXiv1202.0518W}
{Wilde}, M.~M., {Guha}, S., {Tan}, S.-H., \& {Lloyd}, S. 2012,
  \href{https://arxiv.org/abs/1202.0518}{\color{magenta}arXiv},
  \href{https://ui.adsabs.harvard.edu/abs/2012arXiv1202.0518W}{\color{blue}arXiv:1202.0518}.
\newblock \doarXiv{1202.0518}

\bibitem[{{Wright} {et~al.}(2001){Wright}, {Drake}, {Stone}, {Treffers}, \&
  {Werthimer}}]{2001SPIE.4273..173W}
{Wright}, S.~A., {Drake}, F., {Stone}, R.~P., {et~al.} 2001,
  \href{http://dx.doi.org/10.1117/12.435376}{\color{magenta}Proc.~SPIE},
  \href{https://ui.adsabs.harvard.edu/abs/2001SPIE.4273..173W}{\color{blue}4273},
  \href{https://ui.adsabs.harvard.edu/abs/2001SPIE.4273..173W}{\color{blue}173}

\bibitem[{{Wu} {et~al.}(1986){Wu}, {Kimble}, {Hall}, \&
  {Wu}}]{1986PhRvL..57.2520W}
{Wu}, L.-A., {Kimble}, H.~J., {Hall}, J.~L., \& {Wu}, H. 1986,
  \href{http://dx.doi.org/10.1103/PhysRevLett.57.2520}{\color{magenta}\prl},
  \href{https://ui.adsabs.harvard.edu/abs/1986PhRvL..57.2520W}{\color{blue}57},
  \href{https://ui.adsabs.harvard.edu/abs/1986PhRvL..57.2520W}{\color{blue}2520}

\bibitem[{{Yang} {et~al.}(2019){Yang}, {Xu}, {Ren}, {Yin}, {Li}, {Cao}, {Shen},
  {Yong}, {Zhang}, {Liao}, {Pan}, \& {Peng}}]{2019OExpr..2736114Y}
{Yang}, M., {Xu}, F., {Ren}, J.-G., {et~al.} 2019,
  \href{http://dx.doi.org/10.1364/OE.27.036114}{\color{magenta}Optics Express},
  \href{https://ui.adsabs.harvard.edu/abs/2019OExpr..2736114Y}{\color{blue}27},
  \href{https://ui.adsabs.harvard.edu/abs/2019OExpr..2736114Y}{\color{blue}36114}

\bibitem[{{Zhong} {et~al.}(2020){Zhong}, {Wang}, {Deng}, {Chen}, {Peng}, {Luo},
  {Qin}, {Wu}, {Ding}, {Hu}, {Hu}, {Yang}, {Zhang}, {Li}, {Li}, {Jiang}, {Gan},
  {Yang}, {You}, {Wang}, {Li}, {Liu}, {Lu}, \& {Pan}}]{2020Sci...370.1460Z}
{Zhong}, H.-S., {Wang}, H., {Deng}, Y.-H., {et~al.} 2020,
  \href{http://dx.doi.org/10.1126/science.abe8770}{\color{magenta}Science},
  \href{https://ui.adsabs.harvard.edu/abs/2020Sci...370.1460Z}{\color{blue}370},
  \href{https://ui.adsabs.harvard.edu/abs/2020Sci...370.1460Z}{\color{blue}1460}

\bibitem[{{Zhong} {et~al.}(2015){Zhong}, {Hedges}, {Ahlefeldt}, {Bartholomew},
  {Beavan}, {Wittig}, {Longdell}, \& {Sellars}}]{2015Natur.517..177Z}
{Zhong}, M., {Hedges}, M.~P., {Ahlefeldt}, R.~L., {et~al.} 2015,
  \href{http://dx.doi.org/10.1038/nature14025}{\color{magenta}\nat},
  \href{https://ui.adsabs.harvard.edu/abs/2015Natur.517..177Z}{\color{blue}517},
  \href{https://ui.adsabs.harvard.edu/abs/2015Natur.517..177Z}{\color{blue}177}

\end{thebibliography}
\end{document}